\documentclass[10pt]{IEEEtran}

\usepackage{amsmath,graphicx}
\usepackage{amssymb,amstext}
\usepackage{color}  
\usepackage{enumerate} 
\usepackage{subfigure} 
\usepackage{bm}

\usepackage{algorithm}
\usepackage{multirow}
 
\usepackage[noend]{algpseudocode} 
\usepackage{soul}  
\newcommand*{\qed}{\hfill\ensuremath{\square}}%
\newcommand{\betab}{\ensuremath{\boldsymbol{\beta}}}
\newcommand{\beq}{\begin{equation}}
\newcommand{\eeq}{\end{equation}}
\newcommand{\beqa}{\begin{eqnarray}}
\newcommand{\eeqa}{\end{eqnarray}} 
\newcommand{\y}{{\bf y}}
\newcommand{\C}{{\bf C}} 
\newcommand{\bI}{I}
\newcommand{\Real}{\mathbb{R}}

\newcommand{\X}{{\bf X}}
\newcommand{\x}{{\bf x}}

\newcommand{\mub}{{\bm \mu}}
\newcommand{\thetab}{{\bm \theta}}
\newcommand{\Sigmab}{{\bm \Sigma}}

\newcommand{\normalized}{\widetilde}

\newtheorem{Teorema}{\textit{Theorem}}
\newtheorem{Nota}{\textit{Remark}}
\newtheorem{Proposicion}{\textit{Proposition}}
\newtheorem{Corolario}{\em Corollary}

\newcommand{\Na}{\texttt{N1}}

\newcommand{\Nc}{\texttt{N3}}          
\newcommand{\E}{\mathbb{E}}

\usepackage{algpseudocode}% http://ctan.org/pkg/algorithmicx
\algnewcommand\algorithmicinput{\textbf{Input:}}
\algnewcommand\INPUT{\item[\algorithmicinput]}

\algnewcommand\algorithmicoutput{\textbf{Output:}}
\algnewcommand\OUTPUT{\item[\algorithmicoutput]}

\begin{document}

\title{Importance Gaussian Quadrature}

\author{\authorblockN{\authorblockN{V\'ictor Elvira,~\IEEEmembership{Senior Member,~IEEE}, Luca Martino, and Pau Closas,~\IEEEmembership{Senior Member,~IEEE}}\\%  
\thanks{V. Elvira is with the School of Mathematics at the University of Edinburgh (UK), e-mail: victor.elvira@ed.ac.uk; L. Martino is with the Universidad Rey Juan Carlos of Madrid (Spain); P. Closas is with the Department of Electrical and Computer Engineering, Northeastern University, Boston, MA (USA),  e-mail: closas@ece.neu.edu. The work of V\'ictor Elvira was supported by Agence Nationale de la Recherche of France under PISCES project (ANR-17-CE40-0031-01). (Corresponding author: V\'ictor Elvira.) P. Closas was partially supported by the National Science Foundation under Awards CNS-1815349 and ECCS-1845833.}}}

\maketitle

\begin{abstract}
Importance sampling (IS) and numerical integration methods are usually employed for approximating moments of complicated target distributions. In its basic procedure, the IS methodology randomly draws samples from a proposal distribution and weights them accordingly, accounting for the mismatch between the target and proposal.  In this work, we present a general framework of numerical integration techniques inspired by the IS methodology. The framework can also be seen as an incorporation of deterministic rules into IS methods, reducing the error of the estimators by several orders of magnitude in several problems of interest. The proposed approach extends the range of applicability of the Gaussian quadrature rules. For instance, the IS perspective allows us to use Gauss-Hermite rules in problems where the integrand is not involving a  Gaussian distribution, and even more, when the integrand can only be evaluated up to a normalizing constant, as it is usually the case in Bayesian inference. The novel perspective makes use of recent advances on the multiple IS (MIS) and adaptive (AIS) literatures, and incorporates it to a wider numerical integration framework that combines several numerical integration rules that can be iteratively adapted. We analyze the convergence of the algorithms and provide some representative examples showing the superiority of the proposed approach in terms of performance. 
\end{abstract}

\IEEEpeerreviewmaketitle

%\tableofcontents

%%%%%%%%%%%%%%%%%%%%%%%%
%%%%%%%%%%%%%%%%%%%%%%%%
\section{Introduction}
%%%%%%%%%%%%%%%%%%%%%%%%
%%%%%%%%%%%%%%%%%%%%%%%%

The number of applications where it is required to approximate intractable integrals is countless. There is a plethora of approximate methods in the wide of literature in engineering, statistics, and mathematics. These methods are often divided into two main families: the numerical integration (deterministic) methods and the Monte Carlo (random) methods.  

Gaussian quadrature is a family of numerical integration methods based on a deterministic (and optimal, in some sense) choice of weighted points (or nodes) \cite{stoer2013introduction}.\footnote{The term numerical integration is often considered synonym of numerical quadrature, or simply quadrature. Some authors prefer to use the term quadrature for one-dimensional integrands, using the term cubature for higher dimensions \cite{Ballreich2017}. For the sake of brevity, in this paper we will use the term quadrature indistinctly regardless of the dimension.} The approximation is then constructed through a weighted linear combination (according to the weights) of a nonlinear transformation of the points.   This non-linearity, as well as the choice of the nodes and weights, depend on the specific integral to solve. The nodes are deterministically chosen in order to minimize the error in the approximation, which explains the high performance when they can be applied. Thus, when their application is possible, the corresponding algorithms have become benchmark techniques in their fields. As an example in signal processing, Gauss-Hermite rules have been successfully applied in a variety of applications of stochastic filtering, often with remarkable performance \cite{Stano13}. Particularly, the Quadrature Kalman filter (QKF) \cite{Ito-00, Aras-07} and its variants for high-dimensional systems \cite{Clos-10,vila2016uncertainty} showed improved performance over simulation-based methods when the Gaussian assumption on noise statistics holds.
QKF falls in the category of sigma-point Kalman filters, where other variants can be found depending on the deterministic rule used to select and weight the nodes. For instance, one encounters also the popular Unscented Kalman filter (UKF) \cite{Julier00} or the Cubature Kalman filter (CKF) \cite{Aras-09}, both requiring less computational complexity than QKF while degrading its performance in the presence of high nonlinearities \cite{Clos-15}. Moreover, quadrature methods have also been applied in the static framework in a multitude of applications in physics, econometric, and statistics at large \cite{shizgal1981gaussian,gautschi1985gaussian,liu2008use}. However, the application of these methodologies is generally limited to Gaussian noise perturbations in the assumed probabilistic model \cite{sarkka2013bayesian}.

The second family is constituted by the Monte Carlo algorithms, where the nodes are generated randomly (i.e., they are samples) \cite{Robert04,Liu04b}. Arguably, the two main Monte Carlo subfamilies are Markov chain Monte Carlo (MCMC) and importance sampling (IS), and both of them are often used to approximate integrals that involved a specific target distribution. In the former, a Markov chain is constructed in a way that its stationary distribution exists and coincides with the target distribution after a burn-in period. IS simulates samples from a simpler proposal distribution and weights them properly to perform integral approximations. IS provides valid estimators without requiring a burn-in period while enjoys of solid theoretical guarantees such as consistency of the estimators and explicit convergence rates, \cite{owen2013montecarlo,bugallo2017adaptive}.  
Due to their advantages and limitations, in the literature several authors have proposed novel schemes attempting to merge the benefits of  both previous families, e.g.,  including deterministic procedures within the Monte Carlo techniques. This is the case of quasi Monte Carlo methods \cite{caflisch1998monte} and variance reduction methods \cite[Chapter 8]{owen2013montecarlo}.

\noindent{\textbf{Contributions.} 
In this work, we propose a theoretically-grounded framework based on quadrature rules. The IS-based interpretation allow us to propose novel quadrature methods, and pave the way to more sophisticated adaptive mechanisms in very generic settings. We develop the framework by explicitly using the Gauss-Hermite rule (i.e., for Gaussian distributions), but our perspective can be applied to a much wider class of quadrature rules for integration in a variety of sets.  
The basic method on which we develop the framework is referred to as \emph{importance Gauss-Hermite} (IGH) method. We propose a novel estimator, inspired by the self-normalized IS estimator, that can be used when the target distribution can be evaluated only up to a normalizing constant. IGH extends the applicability of {the} Gauss-Hermite rules to a more generic class of integrals which involve other non-Gaussian distributions. This is done by the introduction of the so-called \emph{proposal} density, which is Gaussian in the case of IGH, in a similar manner to the proposal in IS.  We also provide error bounds for the approximations of the integrals in IGH, a related discussion regarding the optimal choice of the proposal function, and a through discussion about the computational complexity.
{Once the IS perspective is introduced, other more sophisticated schemes can be employed, including the use of several proposal pdfs, as in multiple IS (MIS)}
 \cite{elvira2019generalized}, or the adaptation of the proposals as in adaptive IS (AIS) \cite{bugallo2017adaptive}. 
 Recent works have deeply studied the MIS framework, showing for instance that many weighing schemes are possible when more than one proposal are available \cite{elvira2019generalized,elvira2015efficient}. We propose two novel IGH-based schemes with multiple proposals and discuss their performance both from a theoretical point of view and via numerical simulations. Next, we provide some guidelines for the selection and the adaptation of the proposals in IGH. In particular, we propose two novel simple and high-performance adaptive IGH algorithms that are theoretically justified.
 Due to our re-interpretation of quadrature rules from a statistical point of view, we propose statistically inspired mechanisms to adjust the complexity, and a novel metric (named ESS-IGH) for self-assessing {the} new importance quadrature methodology.} 

\noindent{\textbf{Connections to the literature.}}  {In \cite{liu1994note}, the change of measure is proposed in the context of Gauss-Hermite quadrature, using a single Gaussian distribution. This introduced measure, that here we call proposal under our statistical perspective, is set to the Laplace approximation. {The paper considers a unimodal integrand and assumes the maximum to be known.} The relation of this simple change of measure with importance sampling is only mentioned in \cite{pinheiro1995approximations}, although the methodology is not developed. The change of measure is also compared with the Laplace approximation \cite{azevedo1994laplace} (see also \cite{kabaila2019adaptive} for a recent application). In a recent paper \cite{garcia2019gaussian}, the authors apply a change of measure in a more restricted setup (similarly to \cite{liu1994note}), in order to approximate the marginal likelihood with quadrature rules in the context of Gaussian processes. In summary, the methodological power of this change of measure, has not been sufficiently explored in the literature, neither the statistical interpretation of quadrature rules. For instance, the weighted nodes in quadrature methods bear interesting parallelism with importance sampling. A better understanding of these connection will allow in the future for further significant methodological advances.}

\noindent{\textbf{Structure of the paper.}} The rest of the paper is organized as follows. In Section~\ref{sec:problem} we present the problems and briefly discuss importance sampling and numerical integration methods. In Section~\ref{sec_quadIS}, we introduce the importance quadrature framework, particularizing for the case of Gauss-Hermite rules, and introducing the basic IGH method. We discuss the theoretical properties, the choice of the proposal, the computational complexity, and we provide two toy examples and a final discussion where we propose a method for sparse-grids in higher dimensions, and a metric to self-assessed importance quadrature methods. Section~\ref{sec:MIGH} generalizes the IGH for multiple proposals, and we propose two quadrature methods based on two different interpretations coming from the MIS literature. We also discuss the theoretical properties of the methods. Section \ref{sec_adaptive} introduces and adaptive version of IGH, and a discussion about further extensions of the framework. In Section~\ref{sec_simulations} we present three numerical examples:  1) a challenging multimodal target; 2) a signal processing example for inferring the parameters of an exoplanetary system; and 3) a Bayesian machine learning problem for estimating hyperparameters in a Gaussian process (GP). Finally, we conclude the paper with some remarks in Section~\ref{sec_conclusions}.

%%%%%%%%%%%%%%%%%%%%%%%%
%%%%%%%%%%%%%%%%%%%%%%%%
\section{Problem statement and background}
\label{sec:problem}
%%%%%%%%%%%%%%%%%%%%%%%%
%%%%%%%%%%%%%%%%%%%%%%%%

Let us first define a r.v. $\X\in \mathcal{D}\subseteq \mathbb{R}^{d_x}$ with a probability density function (pdf) ${\widetilde \pi}(\x)$. In many applications, the interest lies in computing integrals of the form
\begin{equation}
	\bI = \int_{\mathcal{D}} f(\x) {\widetilde \pi}(\x) d\x,
\label{eq_integral}
\end{equation}
where $f$ can be any integrable function of $\x$ with respect to ${\widetilde \pi}(\x)$. 
Unfortunately, in many practical scenarios, we cannot obtain an analytical solution for Eq. \eqref{eq_integral} and approximated methods need to be used instead.
An illustrative example is the case of Bayesian inference, where the observed data as $\y\in \mathbb{R}^{d_y}$ parametrize the posterior pdf of the unknown vector $\x\in\mathbb{R}^{d_x}$ which is defined as
\begin{equation}
	\normalized{\pi}(\x| \y)
		= \frac{\ell(\y|\x) p_0(\x)}{Z(\y)} \propto \pi(\x|\y)=\ell(\y|\x) p_0(\x),
\label{eq_posterior}
\end{equation}
where $\ell(\y|\x)$ is the likelihood function, $p_0(\x)$ is the prior pdf, and $Z(\y)$ is the normalization factor. This example is even more complicated, since $Z(\y)$ is also unknown, and then $\normalized{\pi}(\x| \y)$ can be evaluated only up to a normalizing constant. {From now on, we remove the dependence on $\y$ to simplify the notation.}

In the following, we review the basics of importance sampling (IS) and deterministic numerical integration with Gaussian distributions.

%%%%%%%%%%%%%%%%%%%
\subsection{Importance sampling (IS)}
\label{sec_is}
%%%%%%%%%%%%%%%%%%%
The basic implementation of IS can be readily understood by first rewriting Eq. \eqref{eq_integral} as 
\begin{align}
	\bI 	&= \int_{\mathcal{D}} f(\x) {\widetilde \pi}(\x) d\x \nonumber \\ 
	&=  \int_{\mathcal{D}} \frac{f(\x) {\widetilde \pi}(\x)}{q(\x)} q(\x)d\x,
\label{eq_integral_IS}
\end{align}
where $q(\x)$ is the so-called \emph{proposal} pdf with non-zero value for all $\x$ where the integrand is non-zero. 
The integral in Eq. \eqref{eq_integral_IS} can be approximated via IS by first simulating a set of $N$ samples $ \{ \x_n \}_{n=1}^N$ from a proposal pdf, $q(\x)$, with heavier tails than $|f(\x)|\pi(\x)$.
Then, each sample is associated an importance weight given by
\begin{equation} 
	w_n= \frac{\pi(\x_n)}{{q(\x_n)}}, \quad n=1,\ldots,N.
\label{is_weights_static}
\end{equation} 
Finally, an unbiased and consistent estimator (with increasing $N$) can be built as
\begin{equation}
	{\widehat{\bI}}_{\textrm{UIS}} = \frac{1}{NZ} \sum_{n=1}^N w_n  f(\x_n),
\label{eq_UIS_estimator}
\end{equation}
which is often denoted as the \emph{unnormalized} importance sampling  (UIS) estimator. 
In many applications, $Z$ is unknown and the UIS cannot be directly applied. 
Instead, using the same samples and weights, the integral in Eq. \eqref{eq_integral} can be approximated with the self-normalized IS (SNIS) estimator as
\begin{equation}
	{\widetilde{\bI}}_{\textrm{SNIS}} = \sum_{n=1}^N \bar w_n  f(\x_n),
\label{eq_partial_estimator_static}
\end{equation}
where $\bar w_i =  \frac{w_i}{\sum_{j=1}^N w_j}$ are the normalized weights. Note that the SNIS estimator can be obtained by plugging the unbiased estimate ${\widehat{Z}}=\frac{1}{N}\sum_{j=1}^N w_j$ instead of $Z$ in Eq. \eqref{eq_UIS_estimator} \cite{Robert04}.
The variance of UIS and SNIS estimators is related to the discrepancy between $\pi(\x)|f(\x)|$ and $q(\x)$, and hence adaptive schemes are usually implemented in order to iteratively improve the efficiency of the method \cite{bugallo2017adaptive}.  
%%%%%%%%%%%%%%%%%%%
\subsection{Numerical Integration based on Gaussian quadrature}
\label{sec_quadmethods}
%%%%%%%%%%%%%%%%%%%
A vast literature in the numerical integration is available, and the specific rules and their justification go beyond the scope of this paper (see for instance in \cite{stoer2013introduction} a review of simple quadrature rules). Here we focus in Gaussian quadrature methods, where a set of weighted nodes are carefully chosen. Common Gaussian quadrature rules are the Gauss-Legendre quadrature for integrals in the bounded domain 
$[-1,1]$ and the Gauss-Hermite (GH) quadrature for integrals involving Gaussian distributions. Moreover, other variants are available, including the Gauss-Kronrod quadrature  and Gauss-Patterson quadrature. {In multidimensional integration, many other rules exist as well, especially with the aim of avoiding an exponential growth of the number of points with the dimension (sparse quadrature rules) \cite[Chapter 8]{owen2013montecarlo}. Some of the most popular approach are the so-called product rule cubature, constructed by directly extending a quadrature rule \cite{engels1980numerical}, or the Smolyak cubature, which is known to be more efficient in the selection of points by exploiting sparsity \cite{smolyak1963quadrature}. The use of sparse grids in multi-dimensional examples allows for computationally efficient integration techniques \cite{heiss2008likelihood}.} For some further details, see Appendix \ref{GQRapp} and Table \ref{tab:QuadratureRules}. In this work, for simplicity, we focus on the GH rule.  However, all the schemes and concepts presented in this work can be easily extended to other Gaussian quadrature rules. Since we mainly focus on the GH rule,
 now we review methods that approximate integrals over Gaussian distributions. Let us consider the integral of the form
\begin{equation}
\bI =  \int_{\mathcal{D}}  h(\x) \mathcal{N}(\x; \mub,\Sigmab) d\x \;,
\label{eq_integral_quadrature}
\end{equation}
\noindent where $\mathcal{N}(\x; \mub,\Sigmab)$ represents a Gaussian pdf with mean $\bm{\mu}$ and covariance $\Sigmab$, and $h$ is a (possibly non-linear) function of the unknown variable $\x$. This integral, which computes a specific moment of a Gaussian distribution $I=\E[h(\x)]$, can be efficiently computed leveraging the aforementioned deterministic rules.
 
Those deterministic methods approximate the integrals with a set of weighted samples/points. 
We refer the interested reader to \cite{golub1969calculation,davis2007methods}. More specifically, the set of deterministic samples and weights are defined as $\mathcal{S} = \{\mathbf{x}_n,v_n \}_{n=1}^N$. Here we focus on the Gauss-Hermite quadrature rules without loss of generality with the aim of being specific, although we point out that the choice of points and weights in $\mathcal{S}$ for approximating the integral in Eq. \eqref{eq_integral} is not unique. The resulting Gauss-Hermite estimator of the integral is given by
\begin{equation}
	\bI \approx \widehat{\bI}_{\textrm{GH}} = \sum_{n=1}^N v_n h(\x_n).
\label{eq_quadrules} 
\end{equation}
{In GH quadrature, the points $\x_n$ are roots of the Hermite polynomial, and the weights $v_n$ are also function of such polynomial (see the last row of Table \ref{tab:QuadratureRules} for an explicit expression and Appendix \ref{GQRapp} for more details). 
It is worth noting that in GH, $N=\alpha^{d_x}$ points are selected, where $\alpha$ corresponds to the number of unique values per dimension (i.e., the resulting points form a multidimensional grid).} 
Therefore, the complexity grows exponentially with the dimension of $\x$ although this issue can be alleviated using other deterministic rules with lower complexity rates such as cubature or unscented rules (requiring $N=2\alpha$ and $N=2\alpha+1$ nodes, respectively) \cite{Clos-15}.
 
In the case of quadrature rules, exact integration in Eq.\eqref{eq_quadrules} occurs when $h(\x)$ is a polynomial of order less or equal than $2 \alpha - 1$. 
Conversely, there is an integration error when the function has a degree higher than $2 \alpha - 1$. For the unidimensional case, $d_x=1$, the error associated to the Gauss-Hermite quadrature rule is related to the remainder of the Taylor expansion of $h(x)$ \cite{gautschi1981survey,Aras-07}
\begin{equation}\label{eq_quadrule_error}
	e = \frac{\alpha! h^{(2\alpha)}(\varepsilon)}{(2\alpha)!},
\end{equation}
\noindent where $ h^{(2\alpha)}(x)$ is the $2\alpha$-th derivative of $h(\cdot)$ and $\varepsilon$ is in the \textit{neighborhood} of $x$. This error analysis can be extended to the multidimensional case, considering that the restriction on the degree should apply per dimension.  
At this point, we would like to notice that \eqref{eq_quadrule_error} can be bounded as
\begin{equation}\label{eq_quadrule_error_2}
	e \leq \frac{\alpha! || h^{(2\alpha)}||_{\infty}}{(2\alpha)!},
\end{equation}
where $||\cdot||_{\infty}$ is the supremum operator. Hence, for any $h(\cdot)$ where the supremum of the  $2\alpha$-th derivative grows slower than $\frac{(2\alpha)!}{\alpha!}$, we can guarantee that the upper bound of the error decreases when we increase the number of quadrature points. Note that in all cases, reducing $||h^{(2\alpha)}||_{\infty}$, implies decreasing the upper bound of the error. 
In Appendix \ref{app_supbound}, we provide a result showing that when $\alpha$ grows, then the bound on the error tends to zero such that $e \rightarrow 0$.

%%%%%%%%%%%%%%%%%%%%%%%%
%%%%%%%%%%%%%%%%%%%%%%%%
\section{Importance Quadrature Schemes}
\label{sec_quadIS}
%%%%%%%%%%%%%%%%%%%%%%%%
%%%%%%%%%%%%%%%%%%%%%%%%
In the following, we develop a novel quadrature-based framework that approximates the integral of Eq. \eqref{eq_integral} for a generic non-Gaussian distribution ${\widetilde \pi}(\x) = \frac{\pi(\x)}{Z}$. To that end, we aim at applying deterministic integration rules under an importance sampling perspective by introducing one or several \emph{proposal} densities. This connection between quadrature methods and IS allows us to develop further non-trivial extensions of quadrature methods, the extension to the case of multiple proposals, the extension of existing adaptive IS procedures, and the development of new adaptive methodologies. We recall that specific importance quadrature methods can be implemented depending on the integration domain $\mathcal{D}$. In the following, and without loss of generality, we focus on $\mathcal{D} = \Real^{d_x}$ and {on} the Gauss-Hermite quadrature rules.

%%%%%%%%%%%%%%%%%%%%%%%%%%%%%%%%%%%%%%%
\subsection{Basic importance Gauss-Hermite (IGH) method}
%%%%%%%%%%%%%%%%%%%%%%%%%%%%%%%%%%%%%%%

Let us rewrite the targeted integral in Eq. \eqref{eq_integral_IS} as
\begin{equation}
	\bI = \int_{\mathcal{D}} h(\x) q(\x) d\x,
\label{eq_integral2}
\end{equation}
where 
\begin{equation}
h(\x)=f(\x) \frac{{\widetilde \pi}(\x)}{q(\x)} \;, \label{eq_integral2_bis}
\end{equation}
and $q(\x)$ is the introduced \emph{proposal} pdf with $q(\x)>0$ for all values where $f(\x) \frac{{\widetilde \pi}(\x)}{q(\x)}$ is non-zero.\footnote{Note that we use the terminology of IS for the proposal $q(\x)$ although the samples are not simulated.} Note that this re-arrangement is the same as the usual IS trick of Eq. \eqref{eq_integral_IS}.
We now choose a Gaussian \emph{proposal} $q(\x) = \mathcal{N}(\x; \bm{\mu},\Sigmab)$, which allows us to re-interpret $I$ as the expectation of $h(\x)$ under the distribution $q(\x)$, as in Eq. \eqref{eq_integral_quadrature}.
 The weighted samples are deterministically chosen with the Gauss-Hermite rules discussed in Section \ref{sec_quadmethods}, reason why we called the method \emph{importance Gauss-Hermite} (IGH) method. Following this double interpretation (from IS and quadrature perspectives), we have an extra degree of freedom in the choice of the parameters of the Gaussian \emph{proposal} pdf $q(\x)= \mathcal{N}(\x; \bm{\mu},\Sigmab)$. 

Let us summarize the basic IGH method in Algorithm \ref{quadIS_algorithm}, which will serve as a basis for further extensions below. In Step 1, $N$ deterministic points, $\{\x_n\}_{n=1}^N$, and their associated quadrature weights $\{v_n\}_{n=1}^N$, are chosen according to the Gauss-Hermite rule. In Step 2, we compute the importance weights according to the standard expression of Eq. \eqref{is_weights_static}. Interestingly, the IGH weights, $\{w'_n\}_{n=1}^N$, are computed as the product of the quadrature and the IS weight, in Eq. \eqref{eq_qIS_weight}.  Note that the weights are multiplied by a factor of $N$, so they can be used at the estimator of $Z$ in Eq. \eqref{eq_Z_estimator_qis}. The unnormalized IGH estimator is given in Eq. \eqref{eq_unnormalized_estimator_qis} in Step 4 (only if $Z$ is known)  while the self-normalized estimator is given in \eqref{eq_selfnormalized_estimator_qis} of Step 5.

\begin{algorithm}

\caption{Basic Importance Gauss-Hermite (IGH) algorithm}\label{alg_qis}
{\small
\begin{algorithmic}[1]
%\Procedure{MyProcedure}{}
\INPUT $N$, $\mub$, $\Sigmab$
\State Select $N$ points $\x_n$ and the associated quadrature weights $v_n$, for $n=1,\dots,N$, considering a Gaussian pdf $q(\x) = \mathcal{N}(\bm{\mu},\Sigmab)$. 
\State Account for the mismatch between $\pi(\x)$ and $q(\x)$ by calculating the importance weights as
\begin{equation}
w_n =  \frac{\pi(\x_n)}{q(\x_n)}, \qquad n=1,\dots,N \;.
\label{eq_is_weights}
\end{equation}
\State Compute the quadrature importance weights as
\begin{equation}
w'_n = w_n v_n N \;,
\label{eq_qIS_weight}
\end{equation}
i.e., the product of the importance weight and the quadrature weight. 
\State The unnormalized estimator is built as
\begin{equation}
\widehat I_{\text{IGH}} = \frac{1}{ZN} \sum_{n=1}^N w'_n f(\x_n)
\label{eq_unnormalized_estimator_qis}
\end{equation}
if $Z$ is known.
\State The self-normalized estimator is built as
\begin{equation}
\widetilde I_{\text{IGH}} = \sum_{n=1}^N \bar w'_n f(\x_n) \;,
\label{eq_selfnormalized_estimator_qis}
\end{equation}
where $\bar w'_n = \frac{w'_n}{\sum_{j=1}^N w'_j}$. The normalizing constant $Z$ can be approximated as 
\begin{equation}
\widehat{Z}_{\text{IGH}} = \frac{1}{N} \sum_{n=1}^N w'_n \;.
\label{eq_Z_estimator_qis}
\end{equation}
\OUTPUT $\{\x_n, w'_n\}_{n=1}^N$
\end{algorithmic}
}
\label{quadIS_algorithm}
\end{algorithm}
{

\subsection{Two toy examples}
\label{sec_toy_examples} 
We present two illustrative toy examples that provide useful insights about the behavior of IGH and the importance of the choice of the proposal, motivating the next sections. 

\subsubsection{\textbf{Toy example 1.} Approximation of the central moments of a modified Nakagami distribution}
\label{sec_toy1}

The goal is to obtain the central moments of a modified Nakagami distribution given by
\begin{equation}
	\normalized{\pi}(x;\mu,\sigma^2,r) = \frac{|x|^r}{Z_{\sigma^2,r} }\exp{\left( -\frac{(x-\mu)^2}{2\sigma^2} \right)} \;,
\label{eq_posterior}
\end{equation}
where $x\in \Real$ and $Z_{\sigma^2,r} = \int |x|^r \exp{\left( -\frac{(x-\mu)^2}{2\sigma^2} \right)} dx$. Note that for some values of the distribution parameters $(\mu,\sigma^2,r)$, $\normalized{\pi}$ is a symmetric version of the Nakagami distribution. We approximate now the first 5 even moments, $p\in\{2,4,6,8,10\}$, with IGH {(note that all odd moments are zero due to the symmetry of the pdf)}. Let us choose the IGH proposal $q(x) = \mathcal{N}(x;\mu,\sigma^2)$, from which we select the  $N$ deterministic weighted points $\{x_n,v_n \}_{n=1}^N$. The unnormalized IGH estimator reduces to 
\begin{align}
\widehat I_{\text{IGH}} &= \frac{1}{Z_{\sigma^2,r}N} \sum_{n=1}^N w'_n h(x_n) \\
&= \frac{1}{Z_{\sigma^2,r}N} \sum_{n=1}^N v_n \frac{\pi(x_n)}{q(x_n)}f(x_n)  \\
&= \frac{1}{Z_{\sigma^2,r}N} \sum_{n=1}^N v_n x^{r+p}_n \;.  
\label{eq_unnormalized_estimator_qis_toy1}
\end{align} 
Note that we have chosen the Gaussian proposal such that the exponential term of the target cancels out with the proposal at the IS weight. Note also that $h(x)= \frac{\pi(x)}{q(x)}f(x)$. According to \eqref{eq_quadrule_error_2}, and since $d_x =1$ and $\alpha = N$, the {error bound} of  \eqref{eq_unnormalized_estimator_qis_toy1} is 
\begin{align}
	|\widehat I - I| \leq \frac{N! || h^{(2N)}||_{\infty}}{(2N)!} =\frac{N! || \left( x^{r+p} \right)^{(2N)}||_{\infty}}{(2N)!} \;.  \label{eq_quadrule_error_nakagami2}
\end{align}
Hence, if $2N>r+p$ then the numerator of Eq. \eqref{eq_quadrule_error_nakagami2} is zero, and the estimator has zero error, i.e., $|\widehat I_{\text{IGH}} - I| = 0$ if the order of the moment satisfies $p\leq 2N-r-1$. Fig. \ref{toy_nakagami_1} shows the relative absolute error of $\widehat I_{\text{IGH}}$ when the number of samples is $N=5$ and the parameter of the target is $r=4$. From Eq. \eqref{eq_quadrule_error_nakagami2}, we know that all moments $p\leq 2N-r-1 = 5$, must be approximated with zero error. Indeed, the figure shows a tiny error of $10^{-15}$ for all $p<5$, due to the finite computer precision. For, $p>5$ however, the error becomes significant. Note that in this case, the upper bound of  Eq. \eqref{eq_quadrule_error_nakagami2} is no longer valid since the bound goes to infinity. {Finally, note that the selected IGH proposal is particularly good for the considered target distribution. The use of the Laplace approximation as proposal would not necessarily achieve a successful performance in this problem (for $N=5$, the relative error is around $10^{-5}$).}

\begin{figure}
\centering
\includegraphics[width=0.35\textwidth]{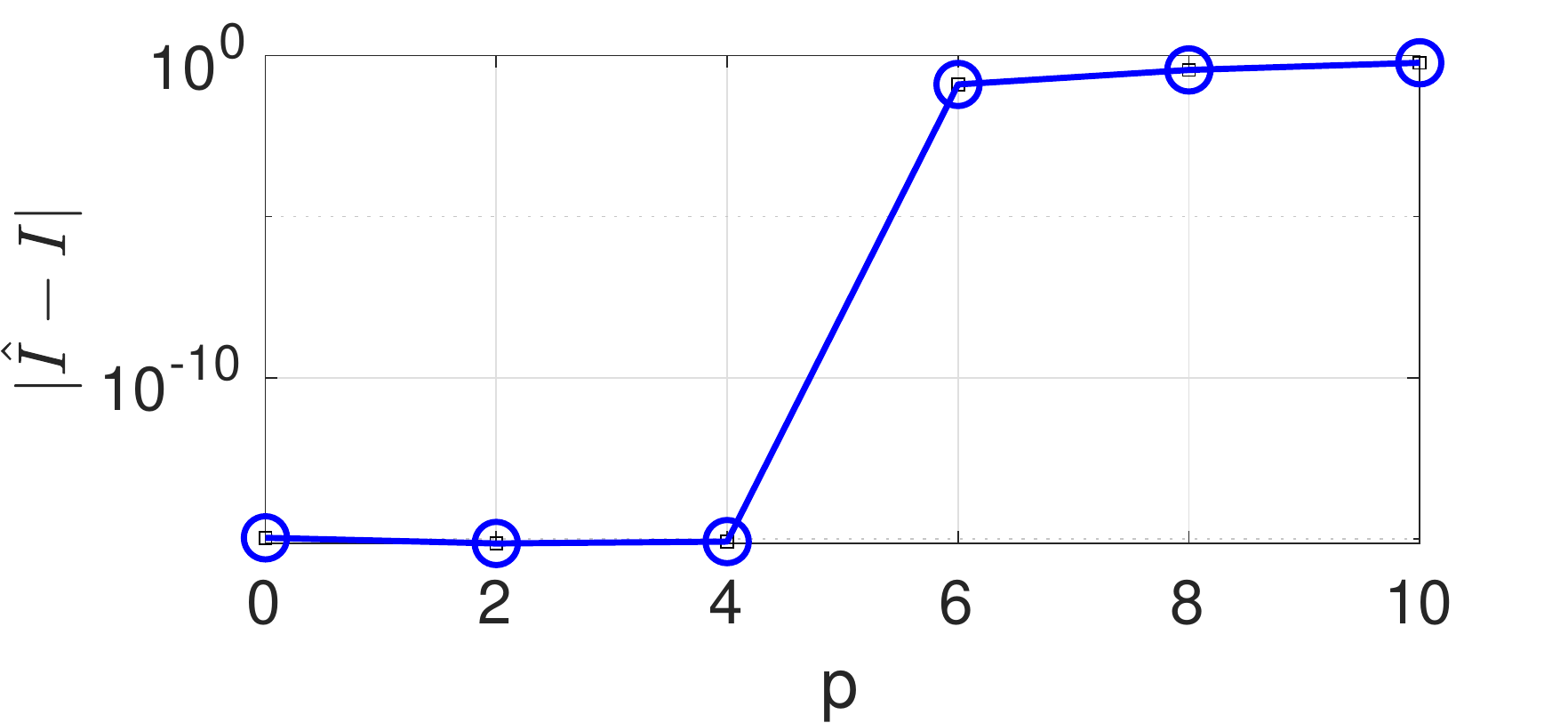}
\caption{\textbf{Toy example 1.} Target $\normalized{\pi}(\x;\mu,\sigma^2,r) = \frac{|x|^r}{Z_{\sigma^2,r} }\exp{\left( -\frac{(x-\mu)^2}{2\sigma^2} \right)}$, {with $r=4$, $\mu=0$, $\sigma=1$}, and $q(x) = \mathcal{N}(x;0,1)$. Function $f(x) = x^p$.}  
\label{toy_nakagami_1}
 
\end{figure}

\subsubsection{\textbf{Toy example 2.} Optimal proposal in IS and IGH} 
{
Let us consider a unidimensional Gaussian target {${\widetilde \pi}(x) = \mathcal{N}(x;1,1)$} and we aim at estimating the mean of the target, i.e., $f(x)=x$, in such a way that we know the solution for this toy problem ($I=0$). 
We apply IS and IGH with the same proposal $q(x) = \mathcal{N}(x;1,\sigma^2)$. We evaluate the performance of the estimators for different values of $\sigma$, using $N=5$ samples/points in both algorithms. Moreover, we use a version of IS where, instead of sampling, we obtain the points using randomized quasi Monte Carlo (QMC) \cite{owen2003quasi}. We name this naive approach \emph{importance QMC} (IQMC). In particular, we obtain the points from the Halton sequence \cite{Niederreiter92} (skipping the first point), and use the Rosenblatt transform so their marginal distribution is the desired normal distribution. We use a randomized version using the Cranley-Patterson rotation \cite[Chapter 17]{owen2013montecarlo}. The results are averaged over $200$ independent runs. Figure \ref{toy_best_sigma}(a) shows the the (mean squared error of the unnormalized estimators in IS (dotted blue), IGH (dashed red), and IQMC (solid black), when $\sigma \in [0.85, 5]$. We also display the squared error of IGH when the Laplace approximation is set as a proposal (dotted gray). Similarly, Figure \ref{toy_best_sigma}(c) shows the (mean) squared error of the self-normalized estimators, and Figure \ref{toy_best_sigma}(c)  displays the (mean) squared error of the normalizing constant estimator. In all figures, the blue circle represents the minimum MSE in IS.

In all IS-based estimators, the minimum MSE is achieved with a $\sigma \in [1, 2]$, but the minimum is not achieved at the same value for the three estimators (see \cite[Chapter 9]{owen2013montecarlo} for a discussion).  
The squared error in the IGH estimators are in general several orders of magnitude below the variance of the corresponding IS estimators. Moreover, the minimum error is achieved for a $\sigma=1$ in the three QIS estimators, which coincides with the standard deviation of the target distribution. Note that IQMC always outperforms IS, but it is still far from the performance of IGH with its optimal proposal. Finally, note that using the Laplace approximation as proposal in IGH provides an adequate performance (but not optimal).  
}

\begin{figure*}
\centering
\subfigure[Mean estimator (unnormalized).]{\includegraphics[width=0.3\textwidth]{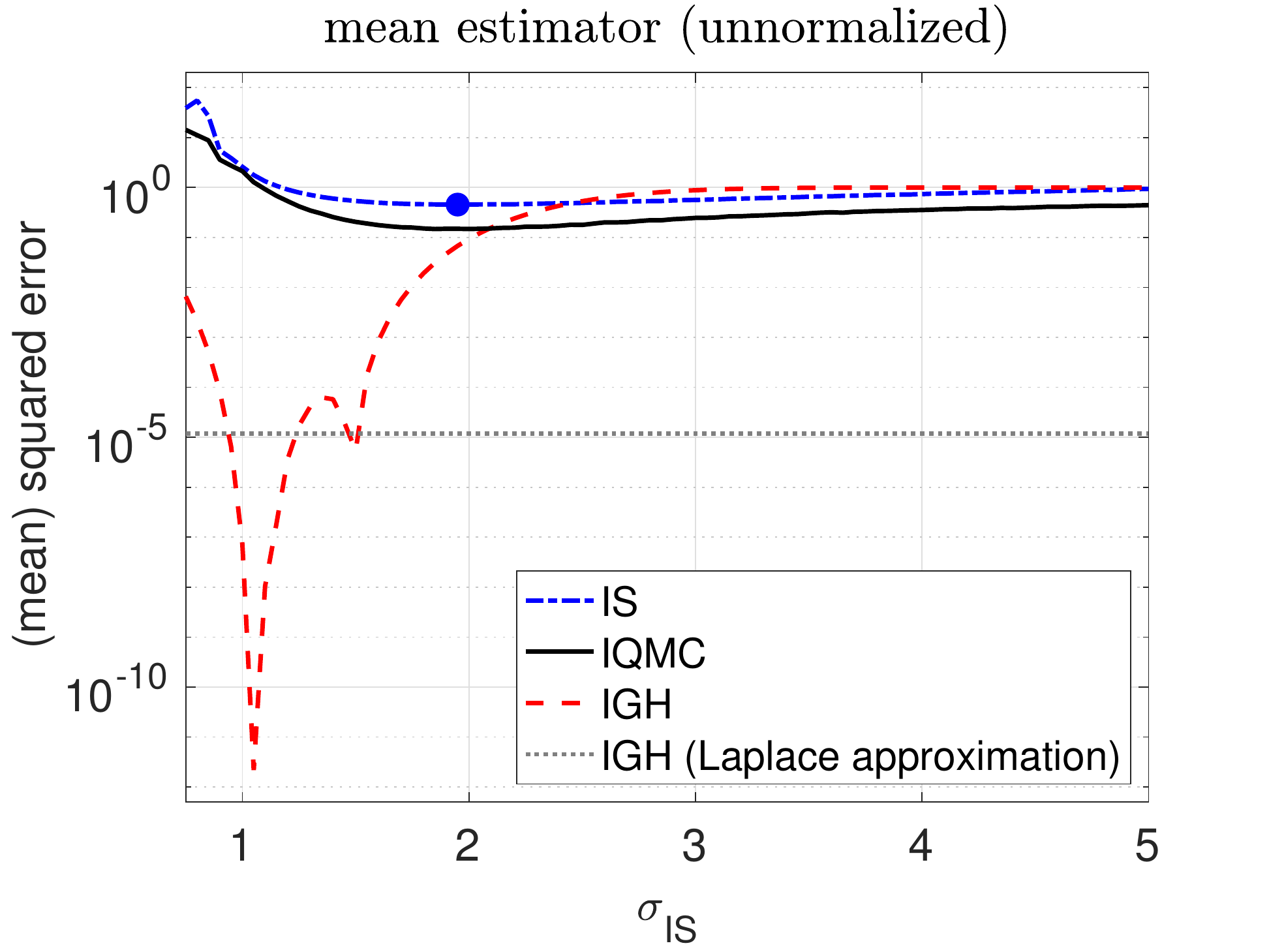}}
\subfigure[Mean estimator (self-normalized).]{\includegraphics[width=0.3\textwidth]{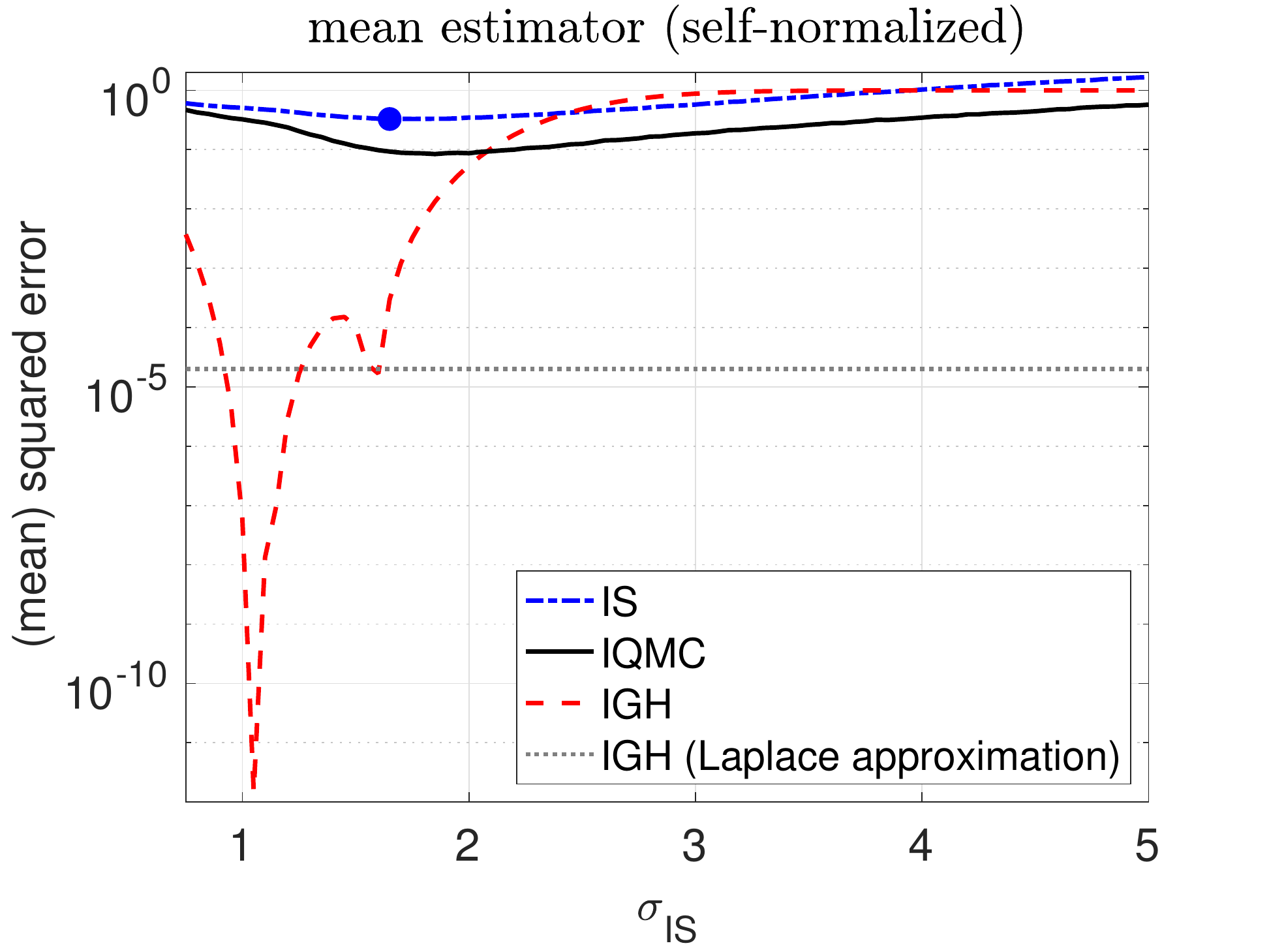}}
\subfigure[Normalizing constant estimator.]{\includegraphics[width=0.3\textwidth]{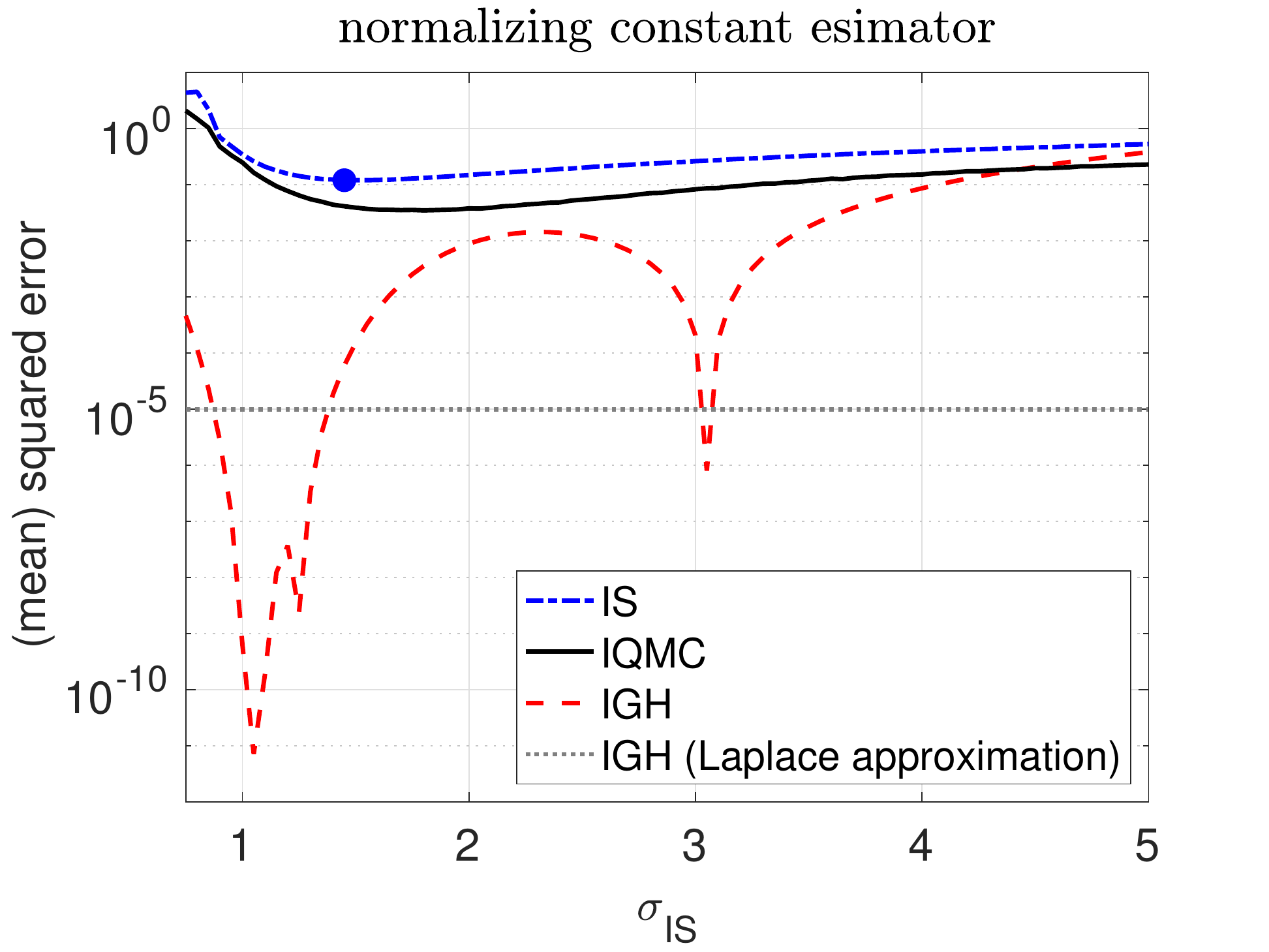}} 
\caption{{\textbf{Toy example 2.} Target ${\widetilde \pi}(x) = \mathcal{N}(x;1,1)$ and $q(x) = \mathcal{N}(x;0,\sigma^2)$, with $\sigma \in \{0.7, 5\}$. We display the (mean) squared error of the following estimators: (a) Unnormalized estimator of the target mean. (b) Self-normalized estimator of the target mean. (c) Estimator of the normalizing constant. The blue circle represents the minimum MSE in IS. The dashed gray line represents the performance of IGH when the proposal is set to the Laplace approximation (hence with a fixed $\sigma$).
}
}
\label{toy_best_sigma}  
\end{figure*}

{In this same setup, now we fix $\sigma = 1.5$ and we approximate the normalizing constant with IS, IQMC, and IGH. Note that the choice of $\sigma$ is particularly good for IS, as shown in Fig. \ref{toy_best_sigma}(c).  In Figure \ref{toy_evol_N}, we show the evolution of the (mean) squared error in IS, IQMC, and IGH for several values of $N\in[3,\;20]$. We also display the IGH with the Laplace approximation as proposal. We see that the convergence rate in this toy example is much faster in IGH than in IS or IQMC.}

{
\subsection{Analysis of the basic IGH and discussion}
\label{sec_discussion}
It is interesting to note that in IS the proposal needs to have heavier tails than the integrand, i.e. $h(\x) = f(\x)\frac{\pi(\x)}{q(\x)}$ must be bounded. In contrast, in the Toy Example $1$, when $p\leq 2N-r-1$, the integrand is $|x|^p\mathcal{N}(x;0,1)$ while the IGH proposal $\mathcal{N}(x;0,1)$ has lighter tails. Let us interpret this from two points of view. On the one hand, regarding Eq. \eqref{eq_quadrule_error_2}, the proposal must be chosen in a way that $h(x)$ is not necessarily bounded, but its $2\alpha$-th derivative is, so the error of the IGH estimator is also bounded. In general, if we aim at a perfect integration, then we need to find a $q(\x)$ such that the $2\alpha$-th derivative of $h(x)$ is zero. On the other hand, in an i.i.d. (random) simulation, the samples are concentrated proportionally to the pdf, while in Gaussian quadrature, from a statistical perspective, the tails are over-represented in terms of nodes (but with an associated weight that is smaller when the distance from the mean to the node grows). For this reason, IGH can still obtain good results with a narrow $q(\x)$.
This suggests that a Laplace approximation of the integrand, as used for instance in \cite{liu1994note}, while providing a good performance in some settings, it is not necessarily the best choice (this is also supported by the two toy examples). The results of IGH from the toy examples $1$ and $2$ are indeed promising, when compared to IS methods. The superior performance comes with some challenges that need to be addressed in order to make IGH a universal methodology that can be used in practical problems. 

\noindent\textbf{Theoretical guarantees.}} Let us first address the convergence of the basic IGH method.
\begin{Teorema}
\label{theorem_conv_IGH}
{The unnormalized, $\widehat I_{\text{IGH}}$, self-normalized $\widetilde I_{\text{IGH}}$, and normalizing constant, $\widehat Z_{\text{IGH}}$, estimators in IGH converge to $I$ when $N\to \infty$.}
\end{Teorema}
\emph{Proof.} See Appendix \ref{appendix_consistency_igh}.\qed
\begin{Nota}
We recall the re-arrangement of \eqref{eq_integral2} is only valid if $q(\x)$ has probability mass for all points where $h(\x) \neq 0$, similarly to what happens in IS. Clearly, this is the case in IGH since $q(\x)$ is Gaussian.
\end{Nota}

The consistency of the estimators ensure the validity of the methodology, but it does not necessarily imply that the approach is efficient for any proposal $q(\x)$. Similarly to IS, the performance of IGH depends on the appropriate choice of a proposal density. Note that the bounds in the approximation error given in Section \ref{sec_quadmethods} apply directly here. We recall that in IS, the optimal proposal that provides a zero-variance UIS estimator is the one proportional to the integrand of the targeted equation as described above in Section \ref{sec_is}. Interestingly, this result is connected to the optimal proposal in IGH.  

\begin{Proposicion} 
{
Let us consider a Gaussian proposal $q(\x;\thetab)$ where $\thetab$ contains both the mean and the covariance matrix, and a function $f$ which is non-negative for all values where $\pi(\x)>0$. Let us suppose that the optimal IS proposal $q^*(\x;\thetab^*) = \frac{f(\x)\pi(\x)}{\int f(\x)\pi(\x) d\x}$ is Gaussian. Then, the same proposal $q^*(\x;\thetab^*)$ used in IGH provides a zero-error unnormalized estimator.} 
\end{Proposicion}
{
\emph{Proof}: By plugging $q^*(\x;\thetab^*)$ in Eq. \eqref{eq_integral2_bis}, then $h^*(\x)=\int f(\x)\pi(\x) d\x=I$, i.e., a constant. Since the Gauss-Hermite rules integrate polynomials perfectly up to order $2\alpha$, the error in this case is zero even with $N=1$ point. \qed
\begin{Nota}
If the optimal IS and IGH optimal proposal does not exist in the parametric form $q(\x;\thetab)$, then the proposal that minimizes the variance of the UIS estimator does not necessarily coincide with the proposal that minimizes the error of the IGH estimator as we show in the second toy example. Note that an extension of the previous proposition can be found in the case that $f$ is non-positive in the support of $\widetilde\pi(\x)$. The case where $f$ takes both positive and negative signs requires the use of multiple proposals (and two samples to obtain a zero-variance IS estimator). More details can be found in \cite[Chapter 13.4]{owen2013montecarlo}
\end{Nota}
}

In real-world problems, it is unlikely that the optimal proposal belongs to the Gaussian family, and hence $h(\x)$ is usually not a constant ({nor} a polynomial) because of the ratio of densities. Therefore, the unnormalized IGH estimator can ensure no error in the estimation of the first $2 \alpha$ terms of the Taylor expansion of $h(\x)$, while integration errors will come from the higher-order terms.  

In the following, we present two toy examples. The first example shows a case where the proposal is chosen in such a way $h(\x)$ is a low-order polynomial, so perfect integration is possible. The second example discusses the best proposal in IS and IGH when perfect integration is not possible, showing that the optimal proposal in IGH is not necessarily the same as in IS.

%%%%%%%%%%%%%%%%%%%
\noindent{\textbf{Computational complexity.}}
%%%%%%%%%%%%%%%%%%%
%
{We first discuss the computational complexity of IGH and related methods for fixed number of points/samples $N$, and then we briefly discuss the selection of $N$. The complexity of deterministic and stochastic integration methods depends on the number of points $N$ at which the target function $h(\cdot)$ needs to be evaluated.  
For instance, in the standard Bayesian inference framework, every point requires the evaluation of all available data, which may be computationally expensive. 
{Recall that the computational cost of drawing a multi-dimensional sample from a Gaussian distribution is ${\mathcal O}(d_x^2)$  \cite{Thomas08}. Additionally, the evaluation of a multivariate Gaussian pdf is ${\mathcal O}(d_x^3)$. 
In Algorithm \ref{alg_qis}, we observe that, since the quadrature points are deterministic, they can be stored and only linear scaling and translation (to adjust for $\bm{\mu}$ and $\Sigmab$) is necessary. As such, the ${\mathcal O}(d_x^2)$ term does not apply in IGH. In contrast, since $h(\cdot)$ involves evaluating $q(\cdot)$, the complexity in IGH is dominated by this operation as $\mathcal{C}_\text{IGH} = {\mathcal O}(N d_x^3)$ under the assumption that the complexity of evaluating $q(\cdot)$ is similar to that of $f(\cdot)$ and ${\widetilde{\pi}}(\cdot)$. Analogously, under Gaussian proposal pdf and same number of points $N$, the IS method has similar complexity $\mathcal{C}_\text{IS} = {\mathcal O}(N d_x^3)$ since the complexity of drawing from $q(\cdot)$ is negligible compared to evaluating from $q(\cdot)$ in the asymptotic analysis (i.e., ${\mathcal O}(d_x^2 + d_x^3) = {\mathcal O}(d_x^3)$). 

IS and Gaussian quadrature algorithms (including the novel IGH framework) require a number of points/samples $N$ that scales exponentially with the dimension, suffering from a similar curse of dimensionality.
In connection to this, some quadrature rules (e.g., Gauss-Hermite) generate a number of points of the form $N=\alpha^{d_x}$, where $\alpha\in\mathbb{N}^+$ is the number of points per dimension (i.e., not all arbitrary choices $N\in\mathbb{N}^+$ are possible). 
This can be cumbersome for some problems, e.g., when it is possible to select a proposal in such a way $h(\x)$ becomes very smooth and one needs a small $N$ (see previous section). In this case, quadrature sparse grids could be used instead \cite{heiss2008likelihood}. However some drawbacks may appear, e.g., in the Smolyak's quadrature rule some weights can be negative, which implies that the numerical approximation can be negative even if the integrand is strictly positive (\cite[Section 3.4]{heiss2008likelihood} and \cite[Chapter 7.8]{owen2013montecarlo}). This can be a problem for importance quadrature techniques when the self-normalized estimator is used, since the normalization of the weights loses our statistical interpretation, and more practically, it can yield to negative estimations of the normalization constant or to even-order moments (see \cite[Section 3.4]{heiss2008likelihood} for more details).
We propose here an alternative to lighten the IGH-based methods by resampling $N'<N$ points with replacement from the pool of $N$ points, with probability equal to its associated quadrature weight. It is easy to show that the new (random) estimator, where the quadrature weights of the resampled points are set to $1/N'$, {is} unbiased w.r.t. to the costly $\widehat I_{\text{IGH}}$ estimator, and converges to it when $N'$ grows (see the example in Section \ref{ex_planets}). Note that this does not reduce the number of points required in order to have a certain level of accuracy, but allows to chose an arbitrary number of points $N'\in\mathbb{N}^+$ where the target will be evaluated (unlike most quadrature rules). Many other similar strategies could be {devised}, although this goes beyond the scope of this paper.}} 

\begin{figure} 
\centering
\includegraphics[width=0.9\columnwidth]{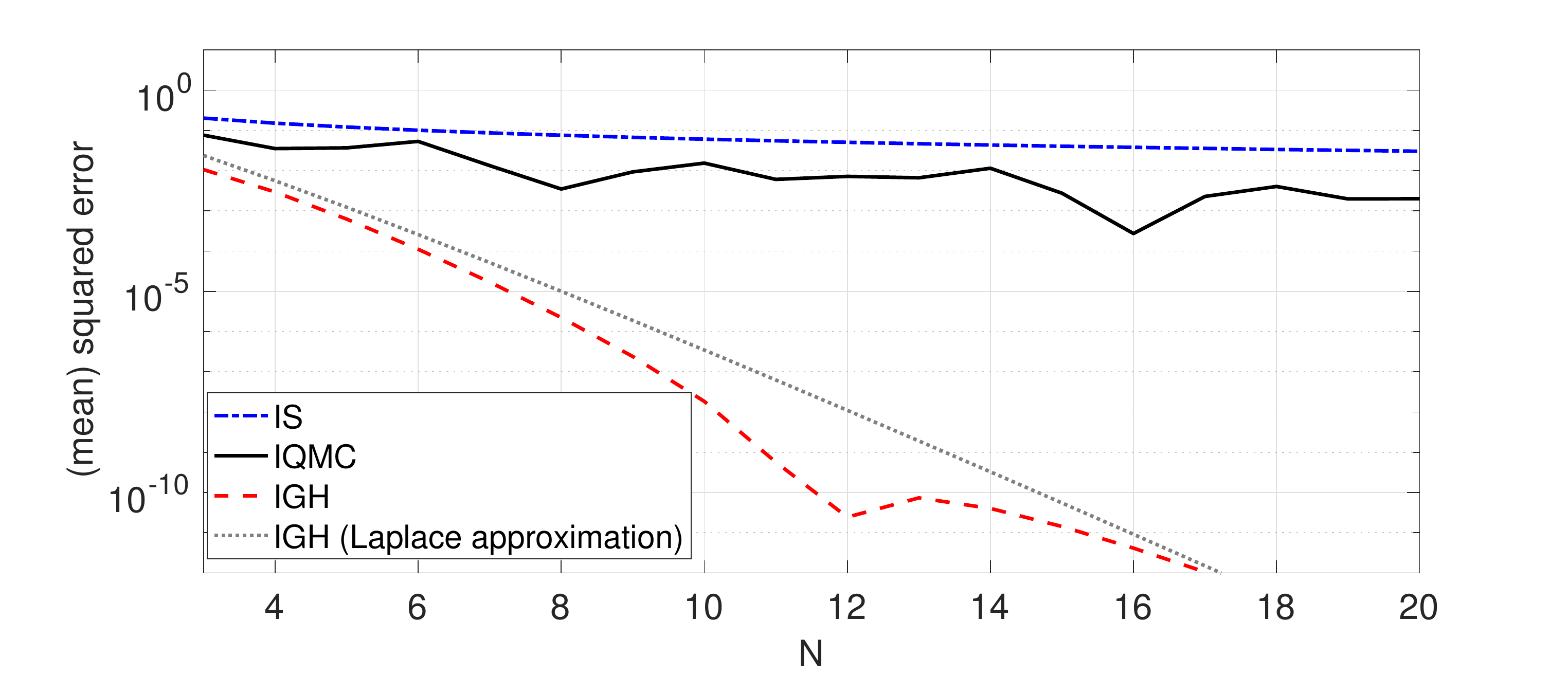}
\caption{Evolution of the (mean) squared error in the approximation of $Z$ in IS and IGH when increasing the number of samples/points $N$.}
\label{toy_evol_N}
\end{figure}

\noindent\textbf{Self-assessed IGH.} Another important issue is the self-assessment of particle-based methods. In IS, the usual measure is an approximation of the effective sample size (ESS). See a discussion about this metric in  \cite{elvira2018rethinking}. We believe that in IGH, another similar metric should be used instead. Following, a recent work about proper metrics in weighted particle-based methods \cite{martino2017effective}, we propose an ESS-like metric for IGH as 
\begin{equation}
\text{ESS-IGH} = \frac{N}{ \frac{N-1}{  \sum_{i\neq j^*} v_i^2 +\left(1-v_{j^*} \right)^2}\left( \sum_{n=1}^N(\bar{w}_n'-{v}_n)^2 \right) + 1}.
\end{equation}
{where $j^*= \arg \min_j v_j$.}
See the derivation and more details in Appendix \ref{appendix_ess}. Note that ESS-IGH fulfills the five desired properties described in \cite{martino2017effective}. 
For instance, the maximum $\text{ESS-IGH}=N$ is only reached in the best scenario when all importance weights $w_n$ are the same (and hence the target is identical to the proposal). Also, the minimum $\text{ESS-IGH}=1$ only occurs in the worst scenario, when only one weight is different from zero, and the {associated} node receives the minimum quadrature point.
Note that, unlike the $\text{ESS}$ in IS, the worst-case scenario happens when the point is the furthest point from the mean of the IGH proposal (which has the smallest $v_j$). In our statistical IGH perspective, this intuition also fits with in this extreme case: not only there is only one effective point, but that the relevant target mass is in the tail of the implicit quadrature proposal (which justifies to receive the minimum $\text{ESS-IGH}=1$). We find this an interesting property, since unlike ESS which is invariant to the node/sample which takes the maximum weight, {the ESS-IGH is more} penalized when the unique non-zero weighted node is further in the tail.   

\noindent\textbf{Automatic IGH.} At this point, we would like a sophisticated method that: $1)$ selects the parameters (mean and covariance) of the proposal density in a way that the integral has minimum error; $2)$ can operate in situations where the target pdf has multiple modes; $3)$ can use more than one {proposal} in order to provide extra flexibility for tackling non-standard distributions; and  $4)$ can adapt to a plethora of complicated problems in an automatic manner. Addressing these challenges is the purpose of the next sections.

%%%%%%%%%%%%%%%%%%%%%%%%%%%%%%%%%%%%%%
%%%%%%%%%%%%%%%%%%%%%%%%%%%%%%%%%%%%%%
%%%%%%%%%%%%%%%%%%%%%%%%%%%%%%%%%%%%%%
\section{Multiple importance Gauss-Hermite method}\label{sec:MIGH}
%%%%%%%%%%%%%%%%%%%%%%%%%%%%%%%%%%%%%%
%%%%%%%%%%%%%%%%%%%%%%%%%%%%%%%%%%%%%%
%%%%%%%%%%%%%%%%%%%%%%%%%%%%%%%%%%%%%%

The novel perspective of the basic IGH method can be extended to the case where it is beneficial to use several proposal densities, $\{q_m(\x)\}_{m=1}^M$. 
In the IS literature, it is widely accepted that using several proposals (or a mixture of them) can improve the performance of the method \cite{Veach95,Owen00,elvira2019generalized}. 
The justification lays on the fact that the efficiency of IS improves when the mismatch between $|f(\x)| {\widetilde \pi}(\x)$ and $q(\x)$ decreases. A mixture of proposals is then more flexible in order to reconstruct the targeted integrand. 

The extension of IGH from single to multiple proposals is not straightforward as we will show below. 
In order to establish the basis of this extension, let us first propose a generic multiple IGH (M-IGH) method in Alg. \ref{alg_migh}. The algorithm receives the parameters of the $M$ proposals, and the number of weighted points per proposal, $N$. {Although $N$ can be different for each proposal, $N_m$, in this paper we will consider that $N_m=N , \forall m$ for simplicity of notation and the explanation.} In Step 1, the $N$ points and associated weights per proposal are chosen. Step 2 computes the importance weights according to some generic weighting scheme $w(\x) = \frac{\pi(\x)}{\varphi_m(\x)}$, {where $\varphi_m(\x)$ is a function that can be different for each proposal} (see below for more details about the choice of $\varphi_m(\x)$). In Step 3, the importance quadrature weights are computed. The unnormalized M-IGH estimator is computed in Eq. \eqref{eq_U_MIGH} of Step 4, and the self-normalized M-IGH estimator in Eq. \eqref{eq_SN_MIGH} of Step 5. Note that again an estimator of the normalizing constant is also available in Eq.~\eqref{Z_migh}.

Similarly to what happens in MIS \cite{elvira2019generalized}, there are several possible re-arrangements of the targeted integral that, introducing the set of $M$ proposals, allow for an integral approximation. In the case of IGH, we can extend the basic re-arrangement in Eqs. \eqref{eq_integral2}-\eqref{eq_integral2_bis} in different ways that will lead to different weighting schemes and interpretations. As we show below, these re-arrangements translate into different implementations of Alg. \ref{alg_migh}, and in particular, in specific choices of the $\varphi_m$ in the weights of Eq. \eqref{eq_qMIS_weight}.

\begin{algorithm}
\caption{Generic Multiple Importance Gauss-Hermite (M-IGH) method}\label{alg_migh}
{\small
\begin{algorithmic}[1]
\INPUT $N$, $\{ \mub_m, \Sigmab_m \}_{m=1}^M$
\State Select $N$ points $\x_{m,n}$ and the associated quadrature weights $v_n$, for $n=1,\dots,N$, associated to each Gaussian pdf $q_m(\x) = \mathcal{N}(\mub_m,\Sigmab_m)$. 
\State Compute the importance weights as
\begin{equation}
w_{m,n} =  \frac{\pi(\x_{m,n})}{\varphi_m(\x_{m,n})}, \;\; m=1,\dots,M; \;\; n=1,\dots,N \;.
\label{eq_mis_weights}
\end{equation}
\State Compute the importance quadrature weights as
\begin{equation}
w'_{m,n} = w_{m,n} v_{n} N \;,
\label{eq_qMIS_weight}
\end{equation}
that is, the product of the importance weight and the quadrature weight.
\State The unnormalized estimator is built as
\begin{equation}
\widehat I_{\text{M-IGH}} = \frac{1}{ZMN} \sum_{m=1}^M  \sum_{n=1}^N w'_{m,n} f(\x_{m,n})
\label{eq_U_MIGH}
\end{equation}
if $Z$ is known.
\State The self-normalized estimator is built as
\begin{equation}
\widetilde I_{\text{M-IGH}} = \sum_{m=1}^M \sum_{n=1}^N\bar w'_{m,n} f(\x_{m,n}) \;,
\label{eq_SN_MIGH}
\end{equation}
where $\bar w'_{m,n} = \frac{w'_{m,n}}{\sum_{i=1}^M\sum_{j=1}^N w'_{i,j}}$. The normalizing constant $Z$ can be approximated as 
\begin{equation}
\widehat{Z}_{\text{M-IGH}} = \frac{1}{MN} \sum_{m=1}^M \sum_{n=1}^N w'_{m,n} \;.
\label{Z_migh}
\end{equation}
\OUTPUT $\{\x_{m,n}, w'_{m,n}\}_{m=1,n=1}^{M,N}$
\end{algorithmic}
}
\end{algorithm}

\subsection{Standard multiple IGH (SM-IGH)}
This approach is a particular case of Alg. \ref{alg_migh}, where the importance weight in Eq. \eqref{eq_mis_weights} for each point $\x_{m,n}$ is computed as  $w_{m,n} =  \frac{\pi(\x_{m,n})}{q_m(\x_{m,n})}$, i.e., $\varphi_m(\x) = q_m(\x)$.
Its derivation follows the re-arrangement of the targeted integral, similar to \eqref{eq_integral2}--\eqref{eq_integral2_bis}, but now involving the $M$ proposal distributions. 
It is possible to rewrite $I$ as 
\begin{align}
	I 	%&= \int  f(\x) \cblue{\widetilde \pi}(\x) d\x \nonumber \\ 
	&=  \frac{1}{M} \sum_{m=1}^M \int  \frac{f(\x) {\widetilde \pi}(\x)}{q_m(\x)} q_m(\x)d\x \label{eq_integral_QN1_a}\\ 
	&=  \frac{1}{M} \sum_{m=1}^M  \int h_m(\x) q_m(\x)d\x \;,
\label{eq_integral_QN1_b}
\end{align}
where $h_m(\x) = \frac{f(\x) {\widetilde \pi}(\x)}{q_m(\x)}$.
Note that it is possible to approximate the $M$ integrals in \eqref{eq_integral_QN1_b} by performing $M$ independent IGH algorithms as in previous section, and the unnormalized estimator of Eq. \eqref{eq_SN_MIGH} is simply the average of the $M$ parallel estimators. The self-normalized estimator of Eq. \eqref{eq_SN_MIGH} however involves the normalization of all $MN$ weights.
Interestingly, the re-arrangement of~\eqref{eq_integral_QN1_a} is inspired in the standard multiple MIS scheme (SM-MIS), denoted $\Na$ scheme in the generalized MIS framework of \cite{elvira2019generalized}. 
For this reason, we denote this algorithm as \emph{standard multiple IGH} (SM-IGH). 
In the SM-MIS scheme, each sample has an importance weight where only the proposal that was used to simulate the sample appears in the denominator.
Note that in \cite{elvira2019generalized} it is shown that the MIS scheme $\Na$ provides a worse performance (largest variance) for the unnormalized estimator in comparison with other MIS schemes (see also \cite[Section 4.1.1.]{elvira2017improving}). 
This poor peformance in MIS is not necessarily translated into a bad performance of the SM-IGH scheme, as we discuss below.
However, both SM-MIS and SM-IGH share the construction of the importance weight as $w_{m,n} =  \frac{\pi(\x_{m,n})}{q_m(\x_{m,n})}$.
The importance weight can be seen as a ratio that measures the mismatch between the target distribution and the denominator of the weight. 
Therefore, in SM-IGH when ${\widetilde \pi}$ has a complicated form that cannot be mimicked with a Gaussian proposal, no matter how many proposals are employed and how their parameters are selected, the mismatch of ${\widetilde \pi}$ with respect to each proposal will be high. 
In other words, a given Gaussian $q_m(\x)$, regardless {of} the choice of its parameters, will  be unable to mimic the target, yielding $h_m(\x)$ very different from a low-order polynomial. {In next section, we propose an alternative scheme to overcome this limitation.}
The following theorem proves the convergence of SM-IGH with $N$.
\begin{Teorema}
\label{theorem_conv_SMIGH}
\emph{The unnormalized and self-normalized SM-IGH estimators converge to $I$ when $N\to \infty$.}
\end{Teorema}
\emph{Proof.} See Appendix \ref{appendix_consistency_smigh}.

%%%%%%%%%%%%%%%%%%%
\subsection{Deterministic mixture IGH (DM-IGH)}
\label{sec_dmigh}
%%%%%%%%%%%%%%%%%%%%%%%

We present a second variant of Alg. \ref{alg_migh} with $\varphi_m(\x) = \frac{1}{M}\sum_{j=1}^M q_j(\x)$, i.e., the same denominator for all samples of all proposals, which is based on an alternative re-arrangement. Let us first define $\psi(\x) \equiv \frac{1}{M} \sum_{m=1}^M  q_m(\x)$, the mixture of all (Gaussian) proposals. The alternative re-arrangement of $I$ that involves $\psi(\x)$ is given by
\begin{align}
	\bI 	
	&= \int  \frac{f(\x) {\widetilde \pi}(\x)}{ \psi(\x) } \psi(\x)  d\x \nonumber \\ 
	&=  \int  \frac{f(\x) {\widetilde \pi}(\x)}{\psi(\x)}\frac{1}{M} \sum_{m=1}^M  q_m(\x)d\x \label{eq_integral_QN3_a}\\ 
	&=  \frac{1}{M} \sum_{m=1}^M  \int h(\x) q_m(\x)d\x \;,
\label{eq_integral_QN3_b}
\end{align}
where now the same function 
\begin{equation}
\label{HminLuca}
    h(\x)= \frac{f(\x) {\widetilde \pi}(\x)}{\psi(\x)}=\frac{f(\x) {\widetilde \pi}(\x)}{\frac{1}{M} \sum_{m=1}^M  q_m(\x)} \;,
\end{equation}
is present in all $M$ integrals.
This re-arrangement is inspired by the deterministic mixture MIS (DM-MIS) scheme, denoted as $\Nc$ in \cite{elvira2019generalized}, where it is proved to provide the smallest variance in the UIS estimator among of all known MIS schemes. Several reasons explain the good behavior of the DM-MIS scheme (see the discussion in \cite[Section 4.1.1]{elvira2017improving}).  
Similarly, in the DM-IGH, the $M$ integrands sharing the same function $h(\x)$ that contains the mixture $\psi(\x)$ with all proposals on its denominator.\footnote{Note that we are forcing the Gaussians in the mixture to be equally weighted, but it would be straightforward to extend the scheme to the case where the mixture is $\psi_{\betab}{(\x)} = \sum_{m=1}^M \beta_m q_m(\x)$ instead.} 
We recall that  ${\widetilde \pi}$ can be skewed, multimodal, or with different tails than a Gaussian, and while the Gaussian restriction in the proposals is limiting, under mild assumptions, any distribution can be approximated by a mixture of Gaussians \cite{Feller66,Sorenson71}.
In the case of DM-IGH, and following similar arguments in Section \ref{sec_discussion}, if the $M$ Gaussians are selected in such a way $h(\x)$ can be approximated by a low-order polynomial, then all $M$ integrals in Eq. \eqref{eq_integral_QN3_b} will be approximated with low error, and the DM-IGH will be accurate. {Note that DM-IGH requires $O(NM^2)$ operations compared to $O(NM)$ operations in SM-IGH}. We now prove the convergence of the DM-IGH method.
\begin{Teorema}
\label{theorem_conv_DMIGH}
\emph{The unnormalized and self-normalized DM-IGH estimators converge to $I$ when $N\to \infty$.}
\end{Teorema}
\emph{Proof.} See Appendix \ref{appendix_consistency_dmigh}.
\begin{Corolario}
As a result of Theorem \ref{theorem_conv_DMIGH}, one can form a partition of proposals and apply the DM-IGH method in each partition, combining then the estimators similarly to the case in MIS \cite{elvira2015efficient,elvira2016heretical,elvira2016overlapping}.
\end{Corolario}

}

\section{Selection and adaptation of the proposal}
\label{sec_adaptive}

The proposed IGH methodology and its variants requires the selection of the mean and covariances of the (potentially multiple) proposal distributions. As in IS, an adequate selection of those parameters is crucial in obtaining the desired results from IGH. In this section, we provide two adaptive extensions to the IGH methodology such that the inference process can be automated and performed adaptively with little practitioner interaction.

\subsection{Adaptive multiple IGH (AM-IGH)}

We propose a first adaptive IGH algorithm that iteratively adapts the proposals through moment matching mechanisms (see \cite{bugallo2017adaptive} for a description of moment-matching strategies in adaptive IS). We describe the new method in Alg. \ref{alg_amigh} naming it as \emph{adaptive multiple IGH} (AM-IGH). The algorithm runs for $T$ iterations\footnote{The term \emph{multiple} comes from the fact that after $T$ iterations, $T$ different proposals have been used (see \cite{CORNUET12} for more details).}, adapting the parameters of the proposal $q^{(t)}(\x) = \mathcal{N}(\mub^{(t)},\Sigmab^{(t)})$ at each iteration $t$.  The importance weights are computed in \eqref{eq_amigh_weights}, where the generic function in the denominator $\varphi^{(t,\tau)}$ is discussed below. Note also that at each iteration, the importance weights corresponding to previous $t-1$ iterations might be also recomputed for a reason that will be apparent below. Then, the quadrature importance weights are computed in Eq. \eqref{eq_amigh_weights_full}, which are then normalized in Eq. \eqref{eq_amigh_weights_full_norm}. Finally, the proposal is adapted through moment matching using the set of all $Nt$ (re)-weighted points. In particular, we match the first and second moments of the target, which allows for the update of the mean and covariance matrix of the proposal.

The generic Alg. \ref{alg_amigh} can be particularized for different choices of the function $\varphi^{(t,\tau)}(\x)$ in the denominator of the weights. One reasonable choice is to use $\varphi^{(t,\tau)}(\x) = q_{\tau}(\x)$, i.e., applying the proposal that was used to choose the points that are being weighted. In this case, it is not necessary to reweight the points of the previous $t-1$ iterations, i.e., only $N$ weight calculations are done at each iteration. Another possible choice is $\varphi^{(t,\tau)}(\x) = \frac{1}{t}\sum_i^{t}q^{(i)}(\x_{n}^{(\tau)} )$. Hence, all the sequence of proposals is used in the mixture of the denominator. However, in order to balance the presence of a proposal in the weight of future points, the past points must also be reweighted to incorporate the future proposals. Therefore, at each iteration $t$, not only the $N$ new points receive a weight, but also the past $N(t-1)$ points need to be reweighted. This has a clear connection with the DM-IGH of Section \ref{sec_dmigh}.  By plugging this choice in Alg. \ref{alg_amigh}, the method has certain parallelism with the celebrated IS-based AMIS algorithm \cite{CORNUET12} that obtains a high performance in a plethora of applications (see \cite{bugallo2017adaptive} for more details). One limitation of this weighting scheme is that the cost in proposal evaluations grows quadratically with $T$ (while it is linear when the choice $\varphi^{(t,\tau)}(\x) = q_{\tau}(\x)$). Another limitation is that the consistency of the AMIS algorithm has not yet been proved (or the lack of it). Recently, a new method for alleviating the computational complexity of AMIS was proposed, also improving the stability of the algorithm \cite{ellaham2019efficient}. The method choses iteratively and automatically a mixture $\varphi^{(t,\tau)}(\x)$ with a reduced  number of components. A similar mechanism can also be used in the proposed AM-IGH framework. 

\begin{algorithm}
\caption{Adaptive Multiple Importance Gauss-Hermite (AM-IGH) method}\label{alg_amigh}
{\small
\begin{algorithmic}[1]
\INPUT $N$, $T$, $\mub^{(1)}$ $\Sigmab^{(1)}$
\For{$t=1, \dots, T$}
\State Select $N$ points $\x_{n}^{(t)}$ and the associated quadrature weights $v_n$, for $n=1,\dots,N$, associated to {the} Gaussian pdf $q^{(t)}(\x) = \mathcal{N}(\mub^{(t)},\Sigmab^{(t)})$. 
\State Compute (and recompute) the (previous) importance weights as
\begin{equation}
w_{n}^{(\tau)} =  \frac{\pi(\x_{n}^{(\tau)})}{\varphi^{(t,\tau)}(\x_n^{(\tau)})},  \;\; n=1,\dots,N; \;\;\; \tau=1,\dots,t.
\label{eq_amigh_weights}
\end{equation}
\State Compute the importance quadrature weights as
\begin{equation}
w_{n}^{'(\tau)} = w_{n}^{(\tau)} v_{n} N,  \;\; n=1,\dots,N;  \;\;\; \tau=1,\dots,t \;,
\label{eq_amigh_weights_full}
\end{equation}
that is, the product of the importance weight and the quadrature weight.
\State Compute the normalized importance weights as
\begin{equation}
\bar{w}_{n}^{'(\tau)} = \frac{w_{n}^{'(\tau)}}{ \sum_{i=1}^t \sum_{k=1}^N w_{k}^{'(i)}} \;,  \;\;\; \tau=1,\dots,t \;.
\label{eq_amigh_weights_full_norm}
\end{equation}
\State Estimate the mean and the covariance of the target with the set of available {weighted} $Nt$ points, and set $\mub^{(t+1)}$ $\Sigmab^{(t+1)}$ to those values.
\EndFor
\end{algorithmic}
}
\end{algorithm}

\subsection{Multiple Population IGH (M-PIGH)}
\label{sec_aigh_beyond}
In many scenarios, the target distribution is multimodal and/or with a shape that cannot be well approximated by a Gaussian distribution. This is well known in the AIS literature, where most of methods employ several proposal densities in order to approximate the target distribution with a mixture the adapted proposals. Examples of the adaptation of multiple proposals in IS can be found in \cite{Cappe04,Cappe08,CORNUET12,martino2015adaptive,elvira2017improving} among many others. 

Here, we propose a second adaptive scheme, called \emph{multiple population IGH} (M-PIGH), whose adaptation relies fully in the deterministic rules re-interpreting the adaptivity mechanism of M-PMC \cite{Cappe08}, an AIS algorithm (hence fully based in Monte Carlo). In summary, the original M-PMC iteratively adapts a mixture proposal of kernels (including the parameters of the kernels and their weight in the mixture) in a stochastic EM-based manner in order to minimize the KL-divergence between the mixture proposal and the target distribution. In M-PIGH, we select quadrature points and weights instead of sampling from the kernels. In order to not over-complicate the novel algorithm with a variable number of points per kernel, we do not adapt the weight of each kernel in the mixture proposal (hence, we do adapt the mean and covariances of the Gaussian kernels). For sake of brevity, we briefly describe the algorithm. M-PIGH adapts a mixture with $M$ equally-weighted Gaussian kernels, that are initialized with some mean and covariance matrices. For $T$ iterations, M-PIGH selects the points and quadrature weights as in IGH, compute the importance weights using the whole mixture in the denominator (implementing the DM-IGH approach) and builds the usual IGH estimators. The means and covariances of next iterations are computed through the Rao-Blackwellized version of the moment matching proposed in \cite{Douc07b} and later implemented in \cite{Cappe08}. A multimodal numerical example is presented in Section \ref{ex_multimodal}, where we compare the proposed M-PIGH and the original M-PMC.
{Finally, Table \label{table_complexity} summarizes computational complexity of all proposed algorithms. Moreover, it displays whether the algorithms use multiple proposals and whether they are adaptive. Note that the AM-IGH algorithm can be implemented with $M=1$ proposal, but also with $M\geq 1$}

\begin{table}[!t]
\centering
{
\caption{Summary of IGH methods in terms of proposal and target evaluations per point.} \label{table_complexity}
\vspace{-0.3cm}
	\begin{center}
	\begin{tabular}{|c|c|c|c|c|c|}
	 \hline
   & standard IGH & SM-IGH  & DM-IGH & AM-IGH & M-PIGH\\
      \hline
      \hline
   proposal eval. & 1  & 1 & $M$ & $TM^*$ & M \\
   \hline
   target eval.  & 1  & 1 & 1 & 1 & 1 \\
   \hline
   multiple prop. &  no & yes & yes & yes$^*$ & yes \\
   \hline
   adaptive &  no & no & no & yes & yes \\
\hline
\end{tabular}
\end{center}
}
\end{table}

%%%%%%%%%%%%%%%%%%%
%%%%%%%%%%%%%%%%%%%
%%%%%%%%%%%%%%%%%%%
\section{Simulation results}
\label{sec_simulations}
%%%%%%%%%%%%%%%%%%%
%%%%%%%%%%%%%%%%%%%
%%%%%%%%%%%%%%%%%%%
 
In the first example, we build a posterior distribution and test the AM-IGH in a challenging signal-processing example. In the second example, we test the M-PIGH in a multimodal scenario. In the third example, we consider a Bayesian machine learning example where the hyperparameters of a Gaussian process are learned.

\subsection{Inference in a exoplanetary model} 
\label{ex_planets}
%%%%%%%%%%%%%%%%%%%%%%%%%%%%%%%%%%%%%%%%%%%%%%%%%%%%%%%%%
In this section, we consider an astrophysics problem that deals with an exoplanetary system \cite{balan2009exofit,feroz2011detecting}.
We consider a simplified model of a Keplerian orbit and the radial velocity of a host star where the observations are given by 
\begin{equation}
y_{r}(t_d)= v +  k \left[ \cos \left(\frac{2\pi}{p} t_d+ \omega \right) + e \cos \left( \omega \right) \right]+ \xi \;,
\label{eq:rv} 
\end{equation}
\noindent where $t_d$, with  $d = 1,\dots ,D$, represent the time instants, $y_{r}(t_d)$ is the $r$-th observation obtained at the $t_d$-th instant, with $r=1,\dots ,R$, $V$ is the mean radial velocity, $k$ is an  amplitude, $p$ is the  period, $\omega$ is  longitude of periastron, $e$ the eccentricity of the orbit and $ \xi\sim \mathcal{N}(0,\sigma_o^2)$ models the variance of the observation noise, $\sigma_o^2$ being known. Note that $t_1,t_2,\dots,t_{D}$ are (known) time instants  where the observations are acquired. In this example, we consider that the five parameters of the system  ($v$, $k$, $p$, $e$, $\omega$) are unknown, i.e., we aim at inferring the random variable ${\bf X}=[V,K,P,E, \Omega]^{\top}$ in dimension $d_x=5$. In this Bayesian inference problem, we consider uniform priors as follows: $p(V)=\mathcal{U}[-15,15]$, $p(K)=\mathcal{U}[0,50]$, $p(P)=\mathcal{U}[0,365]$, $p(E)=\mathcal{U}[0,2\pi]$, and $p(\Omega)=\mathcal{U}[0,1]$.

\begin{figure}
\centering
\includegraphics[width=0.8\columnwidth]{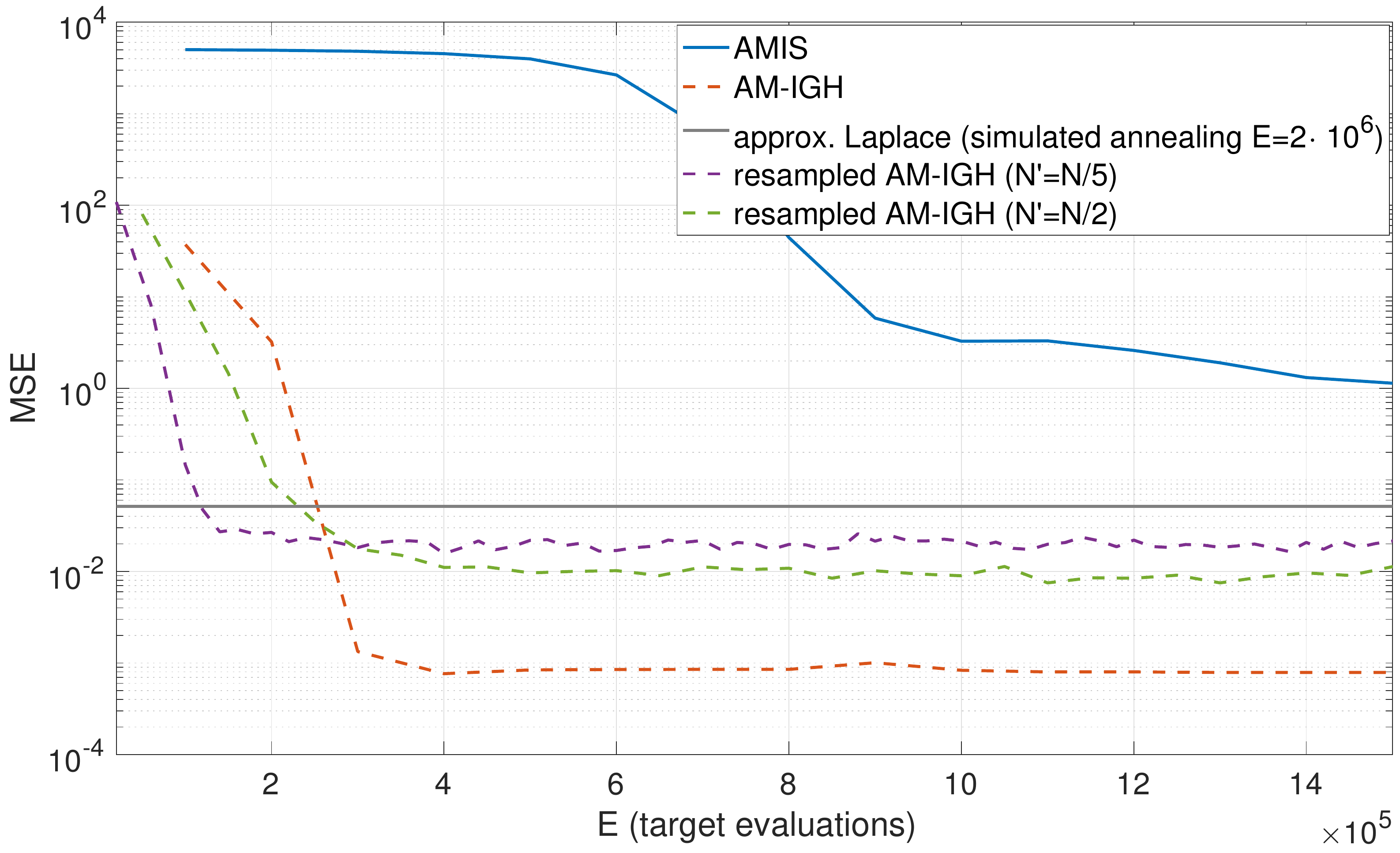}
\caption{{\textbf{Ex. 1.} MSE in the estimate of the mean of the posterior distribution in the exoplanetary model (averaged over the $d_x=5$ dimensions).}}
\label{fig_planetas} 
\end{figure}

For this example, we implement the AM-IGH method with {$N=10^5$ samples/points per iteration, and $T=20$ iterations}. We simulate the model with the values ${\bf X} = [3,2,200,\pi,0.2]^{\top}$, $D=40$ time instants, and $\sigma_o^2=2$ for the observation noise. We made several tests for different values of $R$ and since the results were coherent and did not provide any new insights, we discuss here those with $R=1$.  We  approximate the first moment of the posterior distribution of $\X$ given the set of data.  
{Fig. \ref{fig_planetas} shows the MSE in the estimate of the mean of the posterior distribution building the estimators with the samples at each iteration $t$. {In both AM-IGH and AMIS (for comparison), the means of the Gaussian proposals} have been initialized randomly in the square $[-1.5,6]\times[1,4]\times[100,400]\times[\frac{\pi}{2},2\pi]\times[0.1,0.4]$, and averaged over $100$ independent runs. First, we observe that AM-IGH converges faster to a stable point. Second, AMIS has still not converged to a stable proposal distribution at the last iteration. We recall that the cost of AMIS in proposal evaluations is quadratic with $T$, which becomes a limitation when many iterations are needed to find a good proposal. It is worth noting that the achieved {MSE} of AM-IGH is several orders of magnitude below that of AMIS. Moreover, we have implemented two versions of AM-IGH that used the resampling strategy proposed in Section \ref{sec_discussion}, using $N'=N/2$ and $N'=N/5$ resampled nodes. Interestingly both algorithms converge faster than AM-IGH ($N'=N/2$ is the fastest), but AM-IGH obtains a better performance after few iterations (with $N'=N/2$ being better than $N'=N/5$). The interpretation is simple: in all cases, the best performance is attained after some iterations, and the performance is limited by the number of nodes at the given iteration. We also display the solution given by the Laplace approximation \cite{liu1994note}, finding the mode through a costly simulated annealing \cite{kirkpatrick1983optimization} with $E=2\cdot 10^6$ target evaluations). For unimodal distributions, a good initialization of AM-IGH may be this Laplace approximation, including the use of the Hessian as in \cite{liu1994note}, although it is hard to display a fair comparison since finding the mode is a tough problem when the target is non-concave in the logarithmic domain (as in the considered problem).}

\subsection{Multimodal distribution}
\label{ex_multimodal}
In this example we aim at approximating moments of a multimodal distribution given by the mixture
\begin{equation}
\pi(\x)=\frac{1}{5}\sum_{i=1}^5 \mathcal{N}(\x;{\bm \nu}_i,{\bf \C}_i), \quad \x\in \mathbb{R}^2,
\label{Target1b}
\end{equation}
with the following mean vectors and covariance matrices: ${\bf \nu}_1=[-10, -10]^{\top}$, ${\bm \nu}_2=[0, 16]^{\top}$, ${\bm \nu}_3=[13, 8]^{\top}$, ${\bm \nu}_4=[-9, 7]^{\top}$, ${\bm \nu}_5=[14, -14]^{\top}$, ${\bf \C}_1=[2, \ 0.6; 0.6, \ 1]$, ${\bf \C}_2=[2, \ -0.4; -0.4, \ 2]$, ${\bf \C}_3=[2, \ 0.8; 0.8, \ 2]$, ${\bf \C}_4=[3, \ 0; 0, \ 0.5]$, and ${\bf \C}_5=[2, \ -0.1; -0.1, \ 2]$. 

In this numerical example, due to the multi-modality, we implement M-PIGH, the novel adaptive quadrature method presented in Section \ref{sec_aigh_beyond}. {Unlike in \cite{elvira2019gauss}, here the adaptive mechanism is also based on IGH.}  
\begin{table*}[t]
\caption{\textbf{Ex. 2} MSE in the estimation of the mean and the normalizing constant of the M-PMC (AIS method) and the M-PIGH (novel adaptive quadrature method).}
\label{table_multimodal}
\centering
{\scriptsize
\begin{tabular}{|c|c||c|c|c||c|c|c||c|c|c|}
\cline{3-11}
\multicolumn{2}{c||}{ } & \multicolumn{3}{c||}{$T=5$}  & \multicolumn{3}{c||}{$T=10$}  & \multicolumn{3}{c|}{$T=20$} \\
\cline{3-11}
\multicolumn{2}{c||}{ }  & $\sigma_1 = 1$  & $\sigma_1 = 3$  & $\sigma_1 = 5$  & $\sigma_1 = 1$  & $\sigma_1 = 3$  & $\sigma_1 = 5$  & $\sigma_1 = 1$  & $\sigma_1 = 3$  & $\sigma_1 = 5$  \\
\hline
\multirow{2}{*}{MSE (mean estimate)} &  M-PMC & $46.4$ & $55.6$ & $11.7$  & $67.7$ & $57.9$ & $8.25$ & $72.8$ & $63.1$ & $7.59$ \\   
  & M-PIGH & $18.8$ & $6.94$ & $3.12$ & $9.56$ & $5.13$ & $1.3$  & $8.3$ & $4.21$ & $0.245$ \\ 

  \hline
\multirow{2}{*}{MSE ($Z$ estimate)} &  M-PMC & $1.04$ & $0.681$ & $0.0989$ & $0.824$ & $0.63$ & $0.0299$ & $0.729$ & $0.571$ & $0.026$ \\ 

  &  M-PIGH  & $0.34$ & $0.058$ & $0.034$  & $0.2$ & $0.0385$ & $0.0137$ & $0.141$ & $0.0257$ & $0.00607$ \\ 
\hline
\end{tabular}
}
\vspace*{0.2cm}
\end{table*}
Table \ref{table_multimodal} shows the MSE in the estimation of the mean of the target and the normalizing constant, with both the (stochastic) M-PMC algorithm and the (deterministic) M-PIGH algorithm. In order to compare their behavior, we initialized randomly the location parameters of the kernels  within the $[-4,4]\times[-4,4]$ square, i.e., without covering any modes of the target, in order to better evaluate the adaptivity of the algorithms. For both, M-PMC and M-PIGH we use an adaptive mixture with $M=25$ proposals/kernels, {$N=25$} samples/points per proposal and iteration, for $T\in\{5, 10, 20\}$. {We try three different initializations for the scale parameters of the proposals, with ${\Sigmab_m^{(1)} = \sigma_1^2{\bf I}}$ with $\sigma_1 \in \{ 1, 2, 5 \}$}. The results are averaged over $100$ random initializations.  In all cases, we compare both algorithms with equal number of target evaluations. We see that M-PIGH outperforms M-PMC in all setups, obtaining in some cases an improvement of more than one order of magnitude. For a small scale parameter initialization $\sigma_1 = 1$, both {algorithms} {have trouble improving} their estimate, although M-PIGH is able to significantly improve while M-PMC does not. Larger initial scale parameters benefit both algorithms. We also see that when the number of iterations $T$ is increased, M-PIGH decreases the MSE in a larger factor than the M-PMC: the quadrature rules are not only useful for a better approximation but also for a faster adaptation.

{
%%%%%%%%%%%%%%%%%%%%%%%%%%%%%%%%%%%%%%%%%%%%%%%%%%%%%%%%%
\subsection{Learning Hyperparameters in Gaussian processes with automatic relevance determination}
%%%%%%%%%%%%%%%%%%%%%%%%%%%%%%%%%%%%%%%%%%%%%%%%%%%%%%%%%
Gaussian processes (GPs) are Bayesian state-of-the-art methods for function approximation and regression~\cite{williams2006gaussian}, where selecting the covariance function and learning its hyperparameters is the key to attain significant performance. Here we present an example in the context of estimating the hyperparameters in the \emph{automatic relevance determination (ARD} covariance functions~\cite[Chapter 6]{bishop2006pattern}. 
The observations are $P$ data pairs $\{y_j,{\bf z}_j\}_{j=1}^{P}$, with $y_j\in \mathbb{R}$ and
${\bf z}_j=[z_{j,1},\ldots,z_{j,L}]^{\top}\in \mathbb{R}^{L}$,
 where $L$ is the dimension of the input features. We denote the joint output vector as ${\bf  y}=[y_1,\ldots,y_P]^{\top}$. The goal is to infer an unknown function $f$ which links the variables $y$ and~${\bf z}$~as 
 \begin{equation}
 \label{ModelTrue}
y=f({\bf z})+e,
\end{equation} 
where $e\sim N(e;0,\sigma^2)$. The function $f({\bf z})$ is considered to be a realization of a GP~\cite{williams2006gaussian}, $f({\bf z}) \sim \mathcal{GP}(\mu({\bf z}),\kappa({\bf z},{\bf r}))$, where $\mu({\bf z})=0$, ${\bf z},{\bf r} \in \mathbb{R}^{L}$ with kernel function 
\begin{equation}
\label{EqKernel}
\kappa({\bf z},{\bf r})=\exp\left(-\sum_{\ell=1}^{L}\frac{(z_\ell-r_\ell)^2}{2\delta_\ell^2}\right). 
\end{equation}
The hyper-parameters $\delta_\ell> 0$ corresponding to each input dimension are stacked in ${\bm \delta}=\delta_{1:L}=[\delta_1,\ldots,\delta_L]$.  
 We consider the problem of learning the posterior of all hyper-parameters of the model, given by
 \begin{eqnarray*}
 {\bm \theta}&=&[\theta_{1:L}=\delta_{1:L},\theta_{L+1}=\sigma]=[{\bm \delta}, \sigma] \in \mathbb{R}^{L+1},
\end{eqnarray*}
i.e., all the parameters of the kernel function in Eq.~\eqref{EqKernel} and the standard deviation $\sigma$ of the observation noise.  
We assume a prior $p({\bm \theta})=\prod_{\ell=1}^{L+1}\frac{1}{\theta_\ell^{\beta}}\mathbb{I}_{\theta_\ell}$ where $\beta=1.3$ and $\mathbb{I}_{v}=1$  if $v>0$, and $\mathbb{I}_{v}=0$  if $ v\leq 0$. 
Note that we are focusing on learning the marginal posterior of $p({\bm \theta}|{\bf  y})$, which can be obtained from $p({\bm \theta},{ f}|{\bf  y})$, taking into account that $p({ f}|{\bm  \theta},{\bf  y})$ is tractable  (see \cite{martino2018recycling} for more details about this example).

In Fig. \ref{ResultsGP} we consider the case with $L=3$, so the target is in $\Real^4$, and set a ground truth of ${\bm \delta}^*=[1,3,1]$, $\sigma^*=\frac{1}{2}$  (recall that ${\bm \theta}^*=[{\bm \delta}^*,\sigma^*]$). We have generated $P=500$ pairs of data, $\{y_j,{\bf z}_j\}_{j=1}^{P}$, drawing ${\bf z}_j\sim\mathcal{U}([0,10]^L)$ and ${\bf y}_j$ according to the model in Eq.~\eqref{ModelTrue}. 
We implement the AM-IGH algorithm of Table \ref{alg_amigh} with $\varphi^{(t,\tau)}{(\x)} = q^{(t)}(\x)$, and with $\varphi^{(t,\tau)}{(\x)} = \frac{1}{t}\sum_{j}^t q^{(j)}(\x)$, with $\tau=1,...,t$, which we denote AM-IGH (DM) in the plot. We also compare with the  MC-based method AMIS \cite{CORNUET12}, which incorporates similar weights, {and a QMC version of AMIS, named LAIQMC, where the random sampling is substituted by randomized QMC samples (see the details in the second toy example).} 
We consider also the LAIS algorithm \cite{martino2017layered}, with {$M=1$} proposal, and an IGH version of it, denoted as LA-IGH (also with {$M=1$} proposal). {Moreover, we introduce a variation of LAIS where, the lower layer implements a randomized QMC version (as explained above). We name this version as LAIQMC.} In all LAIS-based algorithms, we use the variant with temporal DM weights (the whole sequence of proposals appear as a mixture in all weights), as described for AM-IGH above. 
More precisely, both LAIS and LA-IGH run a Metropolis-Hastings (MH) chain of length $T$, generating the sequence ${\bm \mu}_1, \ldots ,{\bm \mu}_T$ of location parameters for the IS and IGH methods, respectively. We use a variance $\sigma^2=0.4$ in all algorithms, with the curves being similar for other choices. In all algorithms, we set {$N=625$} samples/nodes per iteration. 
All considered algorithms are iterative, and the comparison is done in terms of target evaluations. 
 We compute the posterior mean (ground truth) with an exhaustive and very costly Monte Carlo approximation so we can compare the methods (see \cite{martino2018recycling} for more details). 
We see that LAIS and LA-IGH exhibit a better performance for a low number of target evaluations (few iterations). The proposed LA-IGH always outperforms LAIS in all the setups we have tested. The proposed AM-IGH and AM-IGH-DM algorithm largely outperform all competitors when the number of iterations is increased. Interestingly, the cheaper version AM-IGH (only one proposal evaluation per sample) is still very competitive w.r.t. the other alternatives, although it provides two orders of magnitude larger error than the AM-IGH-DM.
 \begin{figure}
\centering
\includegraphics[width=0.4\textwidth]{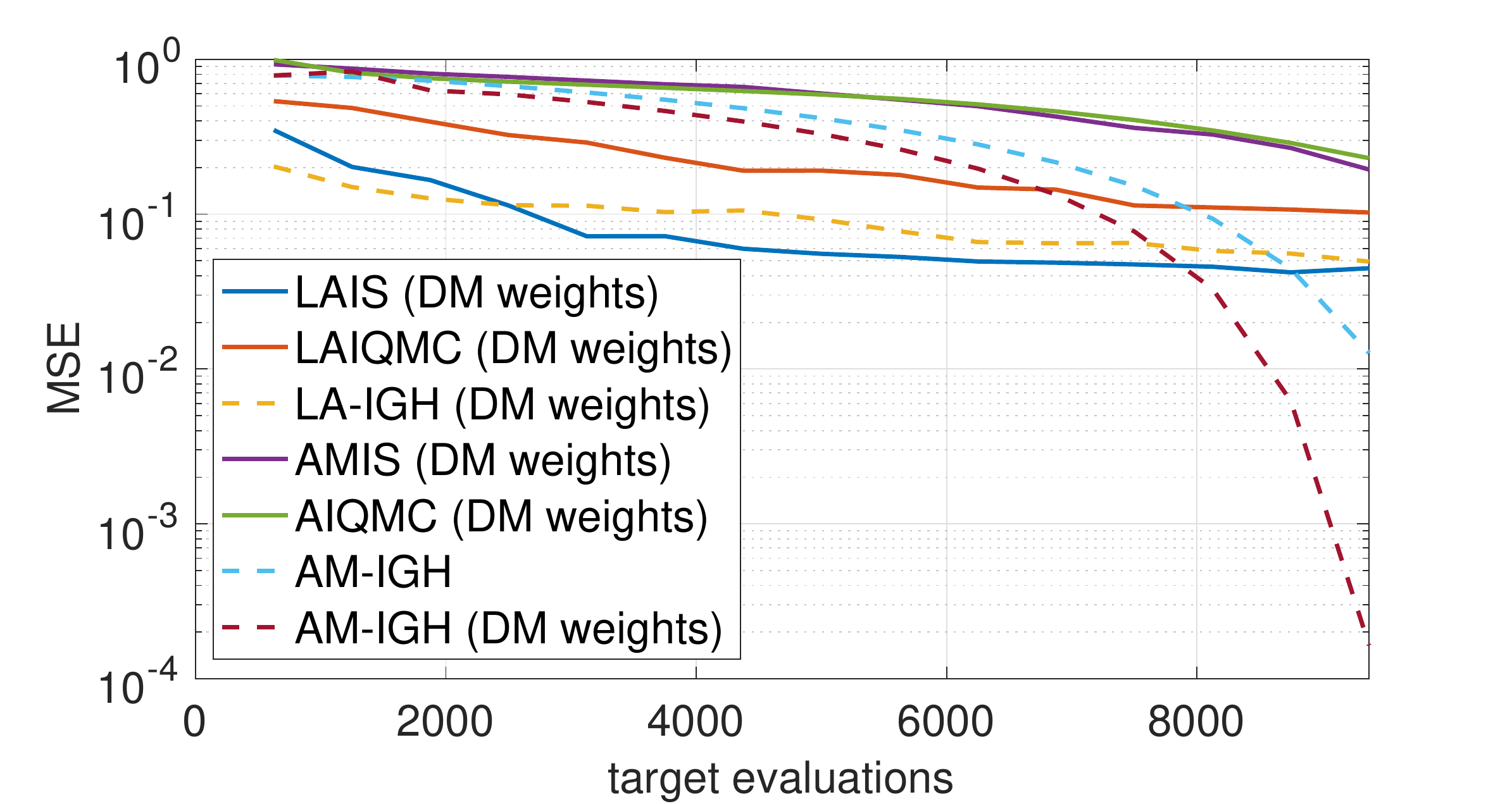} 
\caption{{\textbf{Ex. 3.} Mean squared error of several learning algorithms in the estimation of the mean of the posterior distribution of the hyperparameters of a Gaussian process. All algorithms are compared as a function of the total number of target evaluations (i.e., samples or points) {$E=NMT$}. For each algorithm and configuration, the number of iterations $T$ is set in such a way $E=10^4$.}}
 \label{ResultsGP}
\end{figure} 
}

%%%%%%%%%%%%%%%%%%%
\section{Conclusions}
\label{sec_conclusions}
%%%%%%%%%%%%%%%%%%%

In this paper, we have introduced a generic framework for numerical integration, extending its range of application due to (a) the introduction of a novel importance sampling (IS) perspective, and (b) the incorporation of several ideas from the IS literature. The framework can also be interpreted as an incorporation of deterministic rules into IS methods, reducing the error of the estimators by several orders of magnitude. The potential of the proposed methodology was shown on three numerical examples, as well as two toy examples used in the motivation of the method. This IS perspective allows the use of quadrature rules (in particular, {this work focused on} Gauss-Hermite rules, {although it can be easily applied to other types of Gaussian quadrature rules}) in problems where the integrand does not fulfill the standard requirements in numerical integration. Moreover, the new IS-based framework can also be used when the normalizing constant is unknown, extending its applicability to Bayesian inference.  
{The methodology is completed with a set of extensions, including the use of mixtures of proposals and adaptive approaches to automatically adjust the parameters. Finally, the methodology comes with convergence guarantees and error bounds, which are validated in the discussed examples showing MSE results orders of magnitude below state-of-the-art importance sampling methods.}

\appendices

\section{Proof of Theorem \ref{theorem_conv_IGH}}
\label{appendix_consistency_igh}

First, note that  $\widehat I_{\text{IGH}}$ can be rewritten as in \eqref{eq_quadrules} if $q(\x)$ is non-zero for all $\x$ where ${\widetilde \pi}(\x)$ is non-zero. Then, due to the quadrature arguments reviewed in Section \ref{sec_quadmethods}, $\widehat I_{\text{IGH}}$ converges to $I$. 
The convergence of the self-normalized estimator $\widetilde I_{\text{IGH}}$ is also guaranteed due to similar arguments in IS \cite[Section 3.3.2]{Robert04}. Note that \eqref{eq_selfnormalized_estimator_qis} can be rewritten as $\widetilde I_{\text{IGH}} = \frac{\widehat I_{\text{IGH}} Z}{\widehat{Z}_{\text{IGH}}}$. Both $\widehat I_{\text{IGH}}$, the unnormalized estimator in \eqref{eq_unnormalized_estimator_qis}, and  $\widehat{Z}_{\text{IGH}}$, the normalizing constant estimator in \eqref{eq_Z_estimator_qis}, converge to the desired quantities when $N$ goes to infinity (note that $\widehat Z_{\text{SM-IGH}}$ is a particular case of $\widehat I_{\text{SM-IGH}}$ with $f(\x)=1$). Then, since both the numerator and denominator converge, and since $Z\neq 0$ by construction, we have that $\widetilde I \to I$ when $N$ goes to infinity.\qed

\section{Proof of Theorem \ref{theorem_conv_SMIGH}}
\label{appendix_consistency_smigh}

We first write unnormalized SM-IGH estimator by substituting $\varphi_m(\x)= q_m(\x)$ in Eq. \eqref{eq_mis_weights}:
\begin{align}
\widehat I_{\text{SM-IGH}} &= \frac{1}{ZMN} \sum_{m=1}^M  \sum_{n=1}^N w'_{m,n} f(\x_{m,n}) \nonumber  \\ 
&= \frac{1}{ZMN} \sum_{m=1}^M  \sum_{n=1}^N v_{n} \frac{\pi(\x_{m,n})}{q_m(\x_{m,n})} f(\x_{m,n}) \;.
\end{align}

Due to the properties of the Gauss-Hermite integration, when $N$ goes to infinity,
\begin{align}
\lim_{N\to \infty} \widehat I_{\text{SM-IGH}} &= \frac{1}{ZM} \sum_{m=1}^M  \lim_{N\to \infty} {\frac{1}{N}}\sum_{n=1}^N v_{n} \frac{\pi(\x_{m,n})}{q_m(\x_{m,n})} f(\x_{m,n}) \nonumber\\
&= \frac{1}{MZ} \sum_{m=1}^M  \int \frac{\pi(\x)}{ q_m(\x)} f(\x) q_m(\x)d\x \nonumber \\
&= \frac{1}{Z} \int \pi(\x) f(\x)d\x = I.
\end{align}
Since $\widehat{Z}_{\text{SM-IGH}}$ also converges with $M$ (we recall that it is a particular case of $\widehat I_{\text{SM-IGH}}$ with $f(\x)=1$). Due to the same arguments of Section \ref{sec_discussion}, and the convergence of both $\widehat I_{\text{SM-IGH}}$ and $\widehat{Z}_{\text{SM-IGH}}$ then self-normalized $\widetilde I_{\text{SM-IGH}}$ also converges with $N$. \qed

\begin{table*}[!t]
\centering
\caption{Gaussian Quadrature Rules} \label{tab:QuadratureRules}
\vspace{-0.3cm}
	\begin{center}
	\begin{tabular}{|c|c|c|c|c|}
	 \hline
     {\bf Gaussian Quadrature Rule} & {\bf weighted function $q(x)$} & {\bf Domain $\mathcal{D}$} & {\bf Nodes $x_n$} & {\bf Weights $v_n$}   \\
      \hline
      \hline
      \cline{4-5}
     Legendre  & $q(x) \propto 1$ & $[-1,1]$ &  roots of Legendre polynomials $P_{\alpha}(x)$   & $v_n=\frac{2}{(1-x_n)[P_{\alpha}'(x_n)]^2}$  \\
      \cline{4-5}
 Chebyshev-Gauss &$q(x) \propto \frac{1}{\sqrt{1-x^2}}$ & $(-1,1)$  & $x_n=\cos\left(\frac{2n-1}{2\alpha}\pi\right)$  & $v_n=\frac{\pi}{\alpha}$\\
     \cline{4-5}
      Chebyshev-Gauss-2   & $q(x) \propto \sqrt{1-x^2}$ & $[-1,1]$  & $x_n=\cos\left(\frac{n}{\alpha+1}\pi\right)$ & $v_n=\frac{\pi}{\alpha+1}\sin\left(\frac{n}{\alpha+1}\pi\right)$   \\ 
   \cline{4-5}
      Gauss-Laguerre &  $q(x) \propto \exp(-x)$ & $[0,\infty)$  &  roots of Laguerre polynomials $L_{\alpha}(x)$  & $v_n=\frac{x_n}{(\alpha+1)^2[L_{\alpha+1}(x_n)]^2}$ \\

\cline{4-5}
      Gauss-Hermite &  $q(x) \propto \exp(-x^2)$ & $(-\infty,\infty)$    & roots of Hermite polynomials $H_{\alpha}(x)$ & $v_n=\frac{2^{\alpha-1}\alpha!\sqrt{\pi}}{\alpha^2[H_{\alpha-1}(x_n)]^2}$
      \\   
\hline
\end{tabular}
\end{center}
\end{table*}

\section{Proof of Theorem \ref{theorem_conv_DMIGH}}
\label{appendix_consistency_dmigh}

Let us first write explicitly the unnormalized DM-IGH estimator as 
\begin{align}
\widehat I_{\text{DM-IGH}} &= \frac{1}{ZMN} \sum_{m=1}^M  \sum_{n=1}^N w'_{m,n} f(\x_{m,n})   \nonumber\\ 
&= \frac{1}{ZMN} \sum_{m=1}^M  \sum_{n=1}^N v_{n} \frac{\pi(\x_{m,n})}{\frac{1}{{M}} \sum_{j=1}^{{M}} q_j(\x_{m,n})} f(\x_{m,n}) \nonumber
\end{align}
Again, following quadrature arguments, when $N$ goes to infinity,
\begin{align}
&\lim_{N\to \infty} \widehat I_{\text{DM-IGH}} =  \nonumber\\
&= \frac{1}{ZM} \sum_{m=1}^M  \lim_{N\to \infty} {\frac{1}{N}}\sum_{n=1}^N v_{n} \frac{\pi(\x_{m,n})}{\frac{1}{M} \sum_{j=1}^M q_j(\x_{m,n})} f(\x_{m,n})  \nonumber \\
&= \frac{1}{MZ} \sum_{m=1}^M  \int \frac{\pi(\x)}{\frac{1}{M} \sum_{j=1}^M q_j(\x)} f(\x) q_m(\x)d\x \nonumber \\
&= \frac{1}{Z} \int \frac{\pi(\x)}{\frac{1}{M} \sum_{j=1}^M q_j(\x)} f(\x) \frac{1}{M}\sum_{m=1}^M  q_m(\x)d\x \nonumber \\
&= \frac{1}{Z} \int \pi(\x) f(\x)d\x = I.
\end{align}
Similarly, since $\widehat{Z}_{\text{DM-IGH}}$ also converges with $M$ because of the same reasons as in the IGH and SM-IGH methods, then self-normalized $\widetilde I_{\text{DM-IGH}}$ also converges to $I$ when $N$ goes to infinity. \qed

\section{Bound on the quadrature error}
\label{app_supbound}
We aim at upper bounding the error as in Eq.  \eqref{eq_quadrule_error_2} and showing that asymptotically, as $\alpha \rightarrow \infty$, the error vanishes.
Notice that when the function can be approximated with a polynomial of degree $2\alpha - 1$, the error is zero. Therefore, in the following analysis we are interested in situations where the nonlinearity is such that $p \geq 2\alpha$, where $p$ is the order of the nonlinearity. We make use of useful results regarding bounds of the supremum of a function's derivative \cite{ore1938functions,schaeffer1941inequalities}. 
Let $f_p(x)$ a polynomial of order $p$ such that, in the open interval $(a,b)$, its supremum is bounded by a constant $M_0$, i.e., $\sup |f_p(x)| = M_0$, then the following inequality holds for the first derivative
\begin{equation}
	|f_p^{(1)}(x)| \leq \frac{2 M_0 \; n^2}{b-a} = M_1,
\end{equation}
\noindent and, in general, for the $i$-th derivative we have that
\begin{equation}
	|f_p^{(i)}(x)| \leq K(i,p) \frac{M_0}{(b-a)^i} = M_{i},
\end{equation}
\noindent where
\begin{eqnarray}
	K(i,p) &=& \frac{2^i p^2 (p^2-1) \cdots (p^2 - (i-1))}{1 \cdot 3 \cdot 5 \cdots (2i-1)} \\
    {} 	   &=& \frac{p}{p+i} 2^{2i} \cdot i! \left( \begin{array}{c}
    						p+i \\ 2i \end{array} \right), 
\end{eqnarray}
\noindent and equality only holds for Chebyshev polynomials. We aim at showing that $M_i \geq M_{i+1}$, meaning that the supremum of the derivative $i+1$ is bounded from above by the supremum of the $i$-th derivative. Using the above expressions
$\frac{M_{i+1}}{M_{i}} = \frac{1}{b-a} \frac{p^2 - i}{i(i+1/2)}$, which, for large $i$, tends to $0$ such that $\frac{M_{i+1}}{M_{i}}\leq 1$ is satisfied asymptotically. 
This result is supported by a d'Alembert's ratio test analysis, which states that if the limit of the ratio is such that 
	$\lim_{i\rightarrow \infty} \left| \frac{M_{i+1}}{M_{i}} \right| < 1 \;$, then the series converges absolutely.

%%%%%%%%%%%%%%%%%%%%%%%%%%%%%%%%%%%%%%
\section{Gaussian Quadrature Rules}
%%%%%%%%%%%%%%%%%%%%%%%%%%%%%%%%%%%%%%
\label{GQRapp}
{
For the sake of simplicity and without loss of generality, let us consider $d_x=1$, i.e., $x\in \mathbb{R}$. A quadrature formula $\widehat{I}=\sum_{n=1}^\alpha v_n h(x_n)$ is an approximation of integral of type $I = \int_{\mathcal{D}} h(x) q(x) dx$ in Eq. \eqref{eq_integral2}, i.e.,
\begin{equation}
	I = \int_{\mathcal{D}} h(x) q(x) dx \approx \widehat{I}=\sum_{n=1}^\alpha v_n h(x_n) \;.
\label{eq_integralQuad}
\end{equation}
The function $q(x)$ plays the role of a weighting function (i.e., a density) and it is not required to be normalized, i.e., we only need to assume that $\int_{\mathcal{D}} q(x) dx <\infty$,
i.e., $q(x)$ is an unnormalized density \cite{stoer2013introduction,Ballreich2017}. 
 Given the function $q(x)$, in order to properly select these $2\alpha$ unknown values (all the weights $v_n$'s and all the nodes $x_n$'s), we can consider a nonlinear system of $2\alpha$ equations matching the first $2\alpha$ non-central moments, i.e., 
\begin{eqnarray}
\sum_{n=1}^N v_n x_n^{r}=\int_{\mathcal{D}} x^r q(x) dx, \quad \mbox{ for } \quad r=0,\dots,2\alpha-1 \;,
\end{eqnarray}
where $v_n$'s and $x_n$'s play the role of unknown and the integrals $\int_{\mathcal{D}} x^r q(x) dx$  (i.e., $r$-th moment of $q(x)$) should be a known value. 
Therefore, if the first $2\alpha$ non-central moments  $\int_{\mathcal{D}} x^r q(x) dx$ are available, the non-linear system is well-defined. However, since this system of equations is highly nonlinear, generally the solution is not available \cite{stoer2013introduction,Ballreich2017}. Some specific choices of density $q(x)$ admit a closed-form expression. Table \ref{tab:QuadratureRules} shows some relevant examples.}    

{
%%%%%%%%%%%%%%%%%%%%%%%%%%%%%%%%%%%%%%
\section{ESS-IGH} \label{appendix_ess}
%%%%%%%%%%%%%%%%%%%%%%%%%%%%%%%%%%%%%%
Let us define the Euclidean distance between the two pmfs that define the IGH approximation $\{\bar{w}_n' \}_{n=1}^N$ and the quadrature approximation $\{{v}_n \}_{n=1}^N$ 
\begin{align}
L_2&=\sqrt{\sum_{n=1}^N(\bar{w}_n'-{v}_n)^2} \\
&= \sqrt{\sum_{n=1}^N v_n^2 \left( \frac{w_n}{\sum_{j=1}^N w_j v_j}- 1\right)^2}\\
&= \sqrt{\sum_{n=1}^N v_n^2 \left( \frac{w_n}{\hat Z}- 1\right)^2}.
\label{eq_l2}
\end{align}
When all the importance weights, $w_n$ are equal, $L_2=0$. Following the arguments in \cite{martino2017effective}, the maximum in \eqref{eq_l2} happens when only one importance weight $w_n$ is different from zero. But unlike in IS, here the position of the single non-zero weight plays a role (note that here the nodes are no longer i.i.d. as in IS, and hence they are not exchangeable). Let us denote
$$
j^*= \arg \min_j v_j, 
$$
and hence $v_{j^*}$ is the minimum quadrature weight. Then, the maximum  $L_2$ is
\begin{eqnarray}
L_2^*  &=&  \sqrt{{\sum_{i\neq j^*} (0-v_i)^2} +\left(1-v_{j^*} \right)^2}, \\
&=&  \sqrt{{\sum_{i\neq j^*} v_i^2} +\left(1-v_{j^*} \right)^2}, \label{EqMaxL2}
\end{eqnarray}
i.e., the worst-case is determined by the case where the unique non-zero weight is the one associated to the minimum quadrature weight. In Section \ref{sec_discussion} we given an intuition why this result is relevant. 

Next, we can build a metric ESS-IGH that fulfilled the five desired properties stated in \cite{martino2017effective}, e.g., we would like that ESS-IGH takes its maximum when all the importance weights are the same (which corresponds to the target being identical to the proposal), and its minimum when one extreme point takes the only non-zero weight.  We impose the structure $\text{ESS-IGH} = \frac{1}{aL_2^2+b}$, choosing $a$ and $b$ in such a way  $\text{ESS-IGH}=1$ in the worst scenario ($L_2 =  \sqrt{{\sum_{i\neq j^*} v_i^2} +\left(1-v_{j^*} \right)^2}$), and $\text{ESS-IGH}=N$ in the best case ($L_2 = 0$), which yields $a=\frac{N-1}{N L_2^{*2}}$ and $b=\frac{1}{N}$. Hence,
\begin{align}
\text{ESS-IGH} &= \frac{N}{ \frac{N-1}{{L_2^{2*} }}\left(\sum_{n=1}^N v_n^2 \left( \frac{w_n}{\hat Z}- 1\right)^2 \right) + 1}\\
&= \frac{N}{ \frac{N-1}{{L_2^{2*} }}\left( \sum_{n=1}^N(\bar{w}_n'-{v}_n)^2  \right) + 1}.
\end{align}

}

\vspace{-3mm}
\bibliographystyle{IEEEbib}

\end{document}